\documentclass{emulateapj}

\usepackage{xspace}
\usepackage{graphics,graphicx}
\usepackage{natbib}
\usepackage{multirow}
\usepackage{rotating}

\newcommand{\as}{\mbox{\ensuremath{.\!\!^{\prime\prime}}}}
\newcommand{\asn}{$^{\prime\prime}$\xspace}
\newcommand{\am}{$^{\prime}$\xspace}
\newcommand{\amn}{\mbox{\ensuremath{.\!\!^{\prime}}}}
\newcommand{\nH}{$N_{\rm H}$\xspace}
\newcommand{\PL}{$\Gamma$\xspace}
\newcommand{\Msun}{$M_{\odot}$\xspace}
\newcommand{\Cdof}{$C/dof$\xspace}
\newcommand{\chisq}{$\chi^2$\xspace}
\newcommand{\Xdof}{$\chi^2/dof$\xspace}
\newcommand{\Chandra}{{\it Chandra}\xspace}
\newcommand{\HST}{{\it HST}\xspace}

\newcommand{\lum}{erg s$^{-1}$\xspace}
\newcommand{\flux}{erg s$^{-1}$ cm$^{-2}$\xspace}

\newcommand{\wavdetect}{\texttt{wavdetect}\xspace}
\newcommand{\AEx}{\texttt{AE}\xspace}
\newcommand{\pns}{{\it pns}\xspace}

\newcommand{\lognlogs}{log$N$-log$S$\xspace}
\newcommand{\mdot}{$\dot{m}$\xspace}

\shorttitle{NGC~404 X-ray Point Source Catalog}
\shortauthors{Binder et al.}

\begin{document}

\title{The {\it Chandra} Local Volume Survey: The X-ray Point Source Population of NGC~404}
\author{B. Binder\altaffilmark{1}, 
B. F. Williams\altaffilmark{1}, 
M. Eracleous\altaffilmark{2},
T. J. Gaetz\altaffilmark{3},
A. K. H. Kong\altaffilmark{4},
E. D. Skillman\altaffilmark{5},
D. R. Weisz\altaffilmark{1}
}
\altaffiltext{1}{University of Washington, Department of Astronomy, Box 351580, Seattle, WA 98195}
\altaffiltext{2}{Department of Astronomy \& Astrophysics and Center for Gravitational Wave Physics, The Pennsylvania State University, 525 Davey Lab, University Park, PA 16802}
\altaffiltext{3}{Harvard-Smithsonian Center for Astrophysics, 60 Garden Street Cambridge, MA 02138, USA}
\altaffiltext{4}{Institute of Astronomy and Department of Physics, National Tsing Hua University, Hsinchu 30013, Taiwan}
\altaffiltext{5}{University of Minnesota, Astronomy Department, 116 Church St. SE, Minneapolis, MN 55455}

\begin{abstract}
We present a comprehensive X-ray point source catalog of NGC~404 obtained as part of the \Chandra Local Volume Survey. A new, 97 ks \Chandra ACIS-S observation of NGC~404 was combined with archival observations for a total exposure of $\sim$123 ks. Our survey yields 74 highly significant X-ray point sources and is sensitive to a limiting unabsorbed luminosity of $\sim6\times10^{35}$ \lum in the 0.35-8 keV band. To constrain the nature of each X-ray source, cross-correlations with multi-wavelength data were generated. We searched overlapping \HST observations for optical counterparts to our X-ray detections, but find only two X-ray sources with candidate optical counterparts. We find 21 likely low mass X-ray binaries (LMXBs), although this number is a lower limit due to the difficulties in separating LMXBs from background AGN. The X-ray luminosity functions (XLFs) in both the soft and hard energy bands are presented. The XLFs in the soft band (0.5-2 keV) and the hard band (2-8 keV) have a limiting luminosity at the 90\% completeness limit of $10^{35}$ \lum and $10^{36}$ \lum, respectively, significantly lower than previous X-ray studies of NGC~404. We find the XLFs to be consistent with those of other X-ray populations dominated by LMXBs. However, the number of luminous ($>10^{37}$ \lum) X-ray sources per unit stellar mass in NGC~404 is lower than is observed for other galaxies. The relative lack of luminous XRBs may be due to a population of LMXBs with main sequence companions formed during an epoch of elevated star formation $\sim$0.5 Gyr ago.   
\end{abstract}

\section{Introduction}
The X-ray emission from non-active, early type galaxies is dominated by hot gas \citep{Canizares+87,OSullivan+01} and the X-ray binary (XRB) population. XRBs contain a compact object -- either a neutron star (NS) or black hole (BH) -- accreting from a stellar companion. In low mass XRBs (LMXBs), the companion star is of type A or later, while high mass XRBs (HMXBs) have O or B type companions. Due to the short lifetimes of high mass stars, HMXBs are observed to trace regions of recent star formation \citep{Shty+07}, while the longer-lived lower mass companions of LMXBs follow the underlying, older stellar populations of the host galaxy \citep{Kong+02,Soria+02,Trudolyubov+02}. Thus, the ensemble properties of a galaxy's X-ray point source population, such as the X-ray luminosity function (XLF), should correlate with the star formation history (SFH) and morphology of the host galaxy \citep{Grimm+03,Bel+04,Eracleous+06}.

The \Chandra Local Volume Survey is a deep, volume-limited survey of five nearby galaxies (NGC~55, NGC~300, NGC~404, NGC~2403, and NGC~4214) with overlapping {\it Hubble Space Telescope} (\HST) observations \citep[down to $M_V\sim0$,][]{Dalcanton+09}, designed to test correlations between the shape of the XLF and SFH of the host galaxy. When combined with the already well-studied disks of M~31 \citep{Kong+03} and M~33 \citep{Tullmann+11, Plucinsky+08, Williams+08} available in the literature, these galaxies contain $\sim99$\% of the extragalactic stellar mass and $\sim90$\% of the recent star formation out to $\sim$3.3 Mpc \citep{Tikhonov+03, Maiz+02}. These galaxies additionally span a representative sample of disk galaxies with a range of masses, morphologies, and metallicities. With its excellent angular resolution ($\sim$0\as5) and positional accuracy, \Chandra is the only X-ray telescope capable of separating the X-ray point source populations of nearby galaxies (out to a few Mpc) from diffuse emission. Our collaboration has already examined the X-ray point source population of NGC~300 in \cite{Binder+12}, hereafter referred to as Paper I. 

In this paper, we present the results of a new \Chandra observation of NGC~404. At 3.05 Mpc \citep{Dalcanton+09}, NGC~404 is the closest face-on \citep[$i=11^{\circ}$,][]{delRio+04} S0 galaxy to the Milky Way. Although its optical morphology suggests NGC~404 is a classic ``red and dead'' early-type galaxy, the center appears to have a significant young component \citep{Bouchard+10, Seth+10, Thilker+10,Maoz+98} and may host an intermediate-mass, low-luminosity active nucleus \citep[LLAGN, ][]{Binder+11a,Seth+10,Maoz+05,Eracleous+02}. Here, we focus on the non-nuclear X-ray point source population; the central source is included as part of the X-ray source catalog, but is not discussed in detail in the work \citep[for a discussion of this source, see][]{Binder+11a}.

The organization of this paper is as follows. In Section~\ref{obs} we provide the details of the new observations, the archival data, as well as an outline of our data reduction process and the technique used to generate the X-ray source catalog. Section~\ref{catalog} provides an analysis of the X-ray source catalog, including hardness ratios, timing analysis, spectral fitting for bright sources, and the optical counterpart identification. The XLF and connection to the underlying SFH of NGC~404 is discussed in Section~\ref{sXLF}. We conclude with a discussion and summary in Section~\ref{end}.

\section{Observations and Data Reduction}\label{obs}
\subsection{Observations and Image Alignment}
NGC~404 was observed during the \Chandra X-ray Observatory Cycle 12 on 2010 October 21-22 for 97 ks using the ACIS-S array (Obs. ID 12339), with the nucleus nearly centered on the nominal aim point of the S3 chip. We additionally utilize archival observations: NGC~404 was observed on 1999 December 19 (Obs. ID 870) for $\sim24$ ks and on 2000 August 30 (Obs. ID 384) for $\sim2$ ks, both using the ACIS-S array. Table~\ref{obsdata} summarizes the \Chandra observations used in this work.

\begin{table*}[ht]
\centering
\caption{Summary of \Chandra Observations}
\begin{tabular}{ccccc}
\hline \hline
Obs. ID	& Date		& Exposure (ks)	& R.A. (J2000)		& Decl. (J2000)	\\
(1) 		& (2) 		& (3) 			& (4)				& (5) \\
\hline
384		& 1999 Aug 30	& 1.8				& 01$^h$09$^m$26{\mbox{\ensuremath{.\!\!^{s}}}}90	& +35$^d$43$^m$03{\mbox{\ensuremath{.\!\!^{s}}}}00 \\
870		& 2000 Dec 19	& 24				& 01$^h$09$^m$26{\mbox{\ensuremath{.\!\!^{s}}}}90 	& +35$^d$43$^m$03{\mbox{\ensuremath{.\!\!^{s}}}}40 \\
12239	& 2010 Oct 21-22 & 97			& 01$^h$09$^m$27{\mbox{\ensuremath{.\!\!^{s}}}}00	& +35$^d$43$^m$04{\mbox{\ensuremath{.\!\!^{s}}}}00 \\
\hline\hline
\end{tabular}\label{obsdata}
\end{table*}

Data reduction was performed using CIAO v4.3 and CALDB v4.4.2 using standard reduction procedures. All observations were reprocessed from evt1 level files using the CIAO routine \texttt{acis\_process\_events} with default parameters. The data were corrected for bad pixels/columns, charge transfer inefficiency, and time-dependent gain variations, and we removed the pixel randomization. The event list was then filtered for grades = 0, 2, 3, 4, 6, status = 0, and the pipeline-provided good time intervals. We checked for background flares by creating a background light curve and the source-free event lists on the S3 chip in each observation, and applied a sigma-clipping algorithm to remove time intervals with count rates more than 3$\sigma$ from the mean. Exposure maps and exposure-corrected images were made using \texttt{fluximage} and observations were then merged using \texttt{reproject\_image}. Spectral weights used for the instrument maps were generated assuming a power law spectrum with \PL=1.9 (appropriate for both XRBs and background AGN), and the average foreground column density \nH = 5.13$\times10^{20}$ cm$^{-2}$ \citep{Kalberla+05} was used. 

\begin{figure*}
\centering
\includegraphics[width=1\linewidth,clip=true,trim=0cm 7.5cm 0cm 7cm]{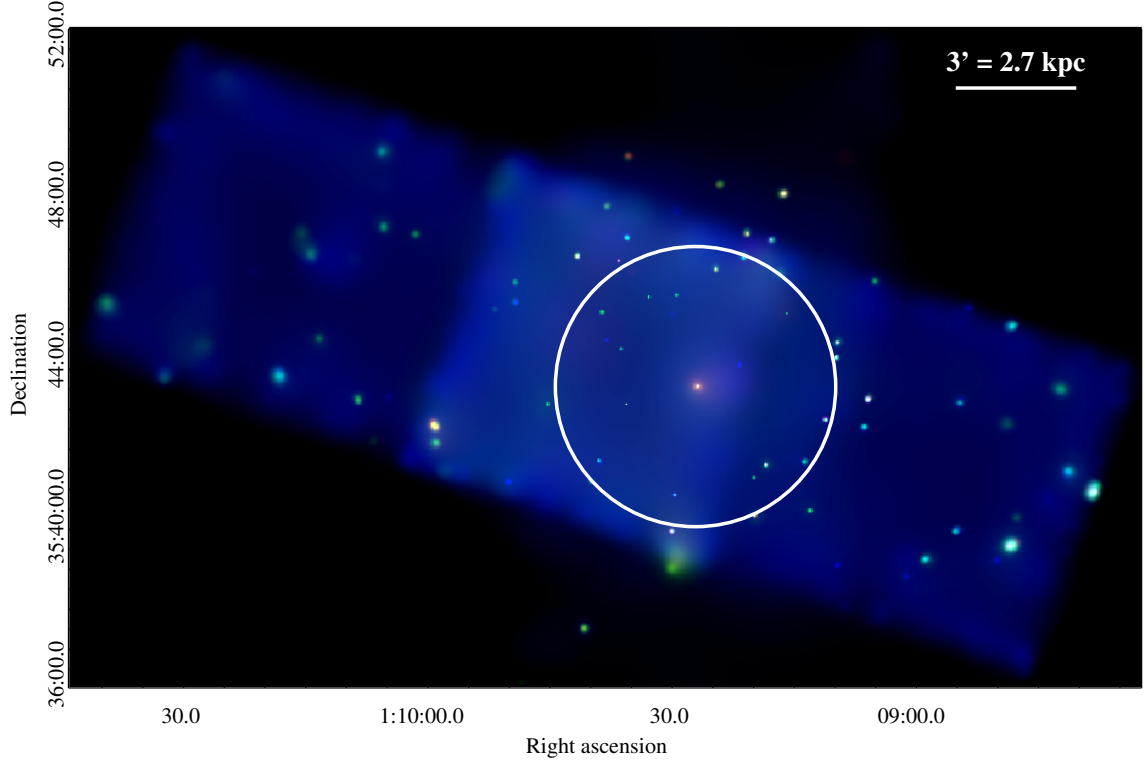} 
\caption{Multicolor exposure-corrected X-ray image of NGC~404. Red represents soft X-ray emission (0.35-1 keV), green represents medium-hard X-rays (1-2 keV), and blue denotes hard X-rays (2-8 keV). The white circle shows the optical extent of the galaxy ($D_{25}$, with a radius of 3\amn5 = 3.1 kpc).}
\label{RGB}
\end{figure*}

\begin{figure*}
\centering
\includegraphics[width=0.65\linewidth,angle=0]{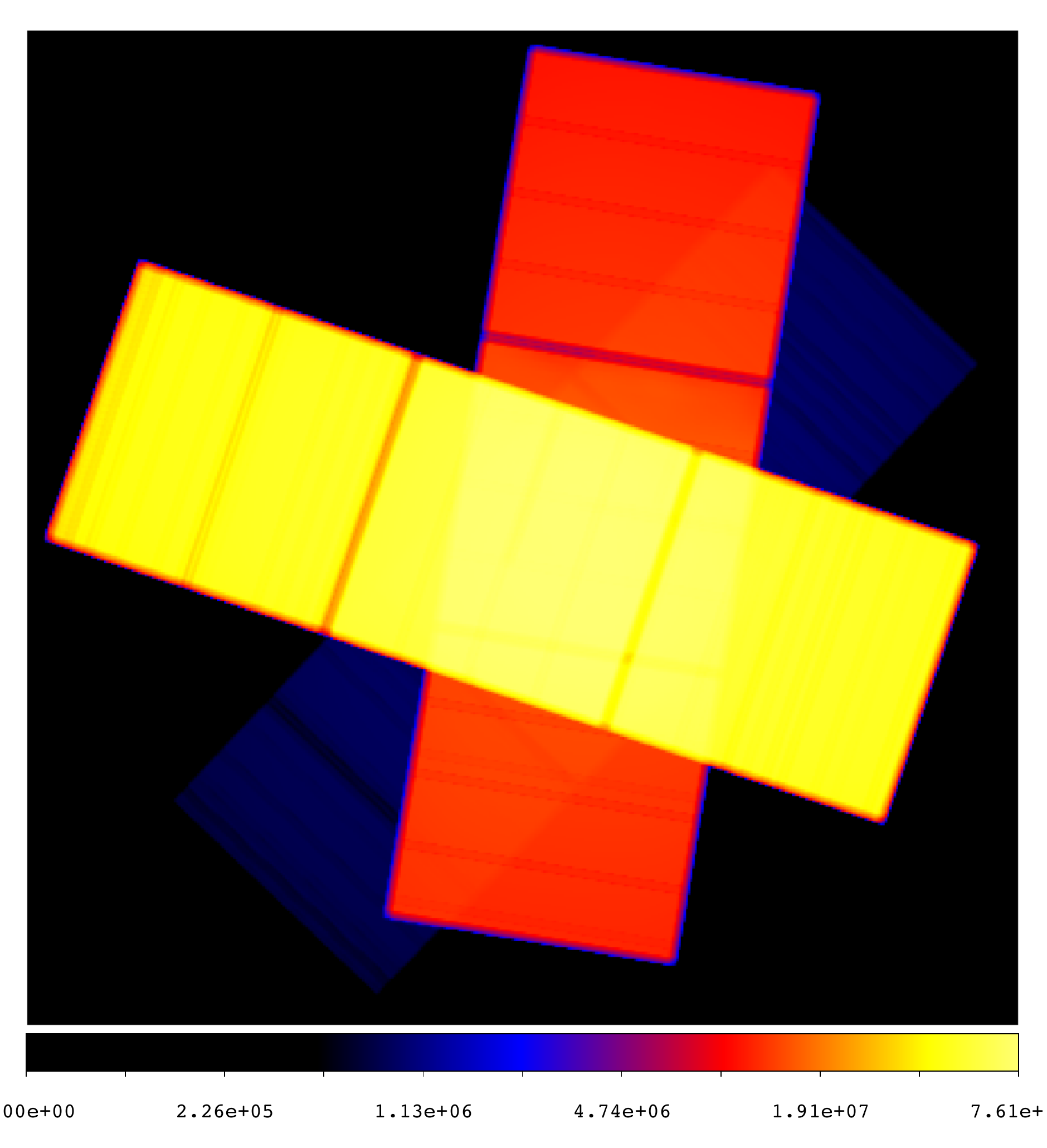} 
\caption{Combined exposure map for all three observations of NGC~404, shown on a logarithmic scale, in units of cm$^2$ s.}
\label{expmap}
\end{figure*}

We created images in the following energy bands (keV): 0.35-8.0, 0.35-1.0, 1.0-2.0, 2.0-8.0 with bin sizes of 1, 2, 3, and 4. A composite X-ray image of NGC~404 is shown in Figure~\ref{RGB}; soft X-rays (0.35-1.0 keV) are displayed in red, medium-hard X-ray emission (1-2 keV) is shown in green, and hard X-rays (2-8 keV) are shown in blue. The optical extent of the galaxy is superimposed in white. The combined exposure map for all three NGC~404 observations is shown in Figure~\ref{expmap}.

\begin{figure*}
\centering
\includegraphics[width=0.85\linewidth,clip=true,trim=1cm 5.5cm 1cm 6cm]{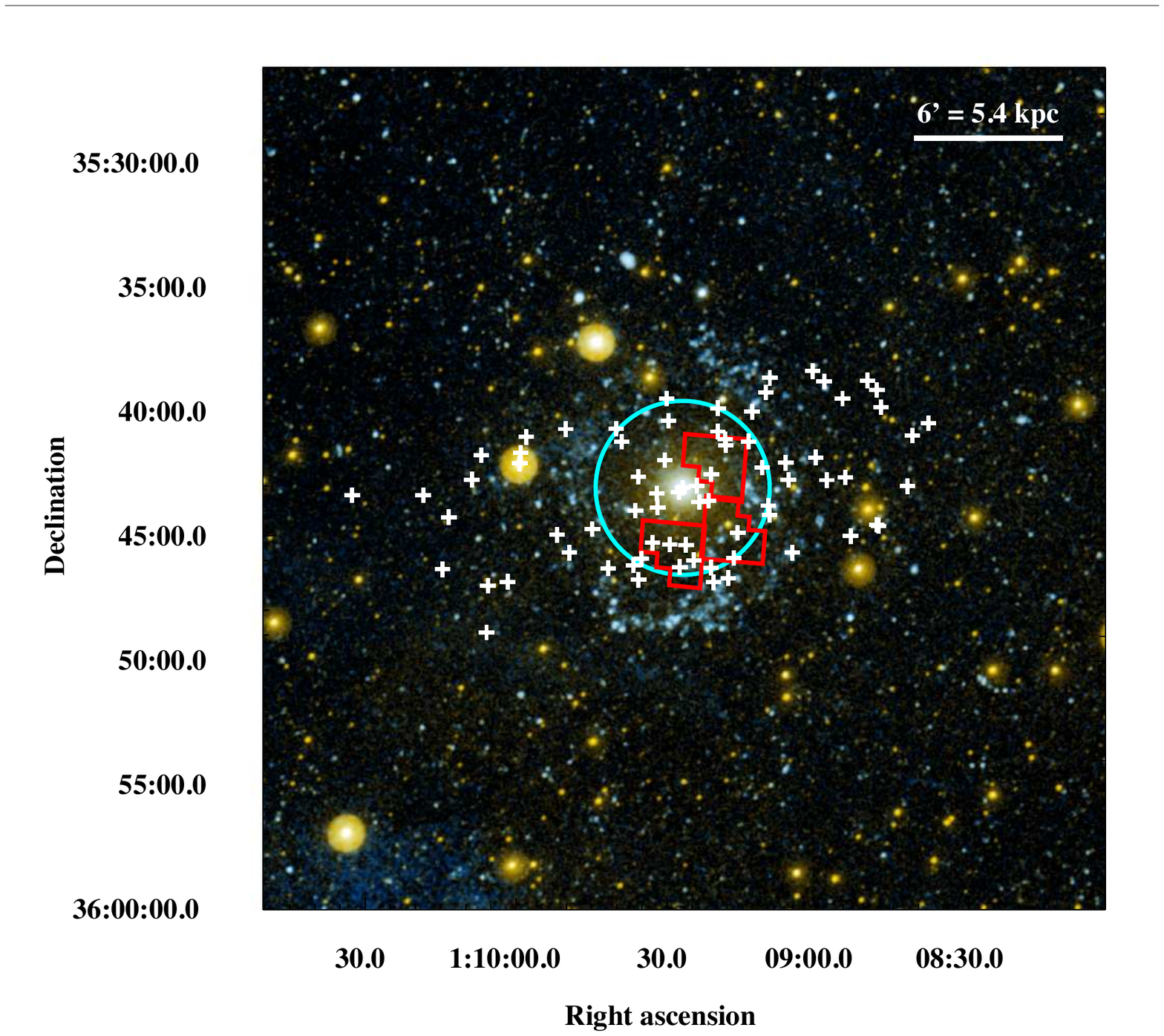} 
\caption{RGB rendered {\it GALEX} image of NGC~404. White crosses indicate the locations of the X-ray sources in the NGC~404 catalog. The optical extent of the galaxy is shown by the cyan circle, and our overlapping \HST fields are outlined in red.}
\label{HST}
\end{figure*}

Three \HST fields were used to search for optical counterparts for each of our X-ray sources. One field (labeled `DEEP') was taken as part of the Advanced Camera for Surveys (ACS) Nearby Galaxy Survey Treasury \citep[ANGST, GO-10915][]{Dalcanton+09}, while the other two shallowered fields (labeled `NE' and `SW') were obtained as part of GO-11986. Details of the \HST data acquisition and data reduction are provided in \cite{Williams+10}. Figure~\ref{HST} shows the locations of our \HST fields and \Chandra X-ray point sources superimposed on a {\it GALEX} image of NGC~404.

To identify candidate optical counterparts to our \Chandra X-ray sources, we place both the \Chandra and \HST fields onto the International Celestial Reference System (ICRS) by finding matches (within $\sim$5\asn) between stars or background galaxies in the Two Micron All Sky Survey (2MASS) Point Source Catalog \citep{Skrutskie+06}. We were able to identify 3-4 bright 2MASS sources per field which matched either a \Chandra X-ray source or optical foreground star or background galaxy. The plate solutions were computed using the IRAF task \texttt{ccmap}, with $rms$ residuals typically less than a few hundredths of an arcsecond in both right ascension and declination. The total alignment error applied to each X-ray source within a given field were computed by summing the \Chandra and \HST $rms$ errors in quadrature. The results of our astrometry are summarized in Table~\ref{alignment}.

\begin{table*}[ht]
\centering
\caption{Alignment of \Chandra and \HST Observations to 2MASS}
\begin{tabular}{cccccc}
\hline \hline
Observatory	& Field				& \# Sources Used	& R. A. $rms^a$		& Decl. $rms^a$	& \# X-ray Sources \\
			&					& for Alignment		&					&				&		\\
(1)			& (2)					& (3)				& (4)					& (5)				& (6)		\\
\hline
\Chandra		& merged				& 3				& 0\as049				& 0\as087			& 74		\\
\HST			& DEEP (Obs. ID 10915)	& 3				& 0\as066 (0\as082)		& 0\as030 (0\as042)	& 3		\\
\HST			& NE (Obs. ID 11986)	& 4				& 0\as041 (0\as064)		& 0\as017 (0\as034)	& 5		\\
\HST			& SW (Obs. ID 11986)	& 4				& 0\as088 (0\as100)		& 0\as001 (0\as087)	& 3		\\
\hline\hline
\end{tabular}\label{alignment}
\tablecomments{$^a$The $rms$ values listed are for each individual \HST field. In parentheses we list the combined $rms$ value for each field, obtained by adding the \Chandra an \HST $rms$ values in quadrature.}
\end{table*}

\subsection{Source Catalog Creation}
We employed the same iterative source detection strategy as in Paper I \citep[see also][]{Tullmann+11}. We present a summary of the approach here; the reader is referred to Paper I and \cite{Tullmann+11} for details. Additionally, we provide a direct comparison of this iterative source detection strategy to the more traditional approach of using the CIAO task \wavdetect \citep{Freeman+02} in the Appendix.

The CIAO task \wavdetect \citep{Freeman+02} was first used to create a list of potential source candidates to be used as input to \texttt{ACIS-Extract} \citep[\AEx;][]{Broos+10}. \AEx is a source extraction and characterization tool which was used to determine various source properties (source and background count rates, detection significances, fluxes, etc.) and to generate light curves and spectra (with appropriate response matrices). We use the Poisson probability of not being a source ({\it prob\_no\_source}; hereafter \pns) provided by \AEx as our threshold criteria when constructing the NGC~404 point source catalog.

For each input source, \AEx creates background regions by first removing all sources from the data and then searching for the smallest circular region around each source that encompasses at least the minimum number of background counts specified by the user. We require each background area to contain at least 50 counts \citep[][found that changing the number of background counts required from 50 to 100 does not significantly affect source properties]{Tullmann+11}. These circular background regions are additionally the regions used to extract background spectra. To extract the X-ray spectrum for each source, \AEx constructs a polygon region that approximates the 90\% contour of the \Chandra-ACIS PSF at the location of the source. The CIAO tools \texttt{mkacisrmf} and \texttt{mkarf} are used to create appropriate response and ancillary response matrices for each source.

The initial source list generated by \wavdetect was deliberately made to include many more sources (762) than we anticipated being statistically significant. We then applied the iterative procedure described in Paper I and \cite{Tullmann+11} to remove false sources. Our final iteration required that a candidate X-ray source have \pns $<4\times10^{-6}$ \citep[see also ][]{Nandra+05,Georgakakis+08,Tullmann+11} in any of the following energy bands: 0.5-8, 0.5-2, 2-8, 0.5-1, 1-2, 2-4, 4-8, 0.35-1, or 0.35-8 keV.

The final CLVS source catalog for NGC~404 contains 74 sources\footnote{The source catalog is available in FITS format at \url{http://www.astro.washington.edu/users/bbinder/CLVS/}}. Their positions and properties, such as detection significance (\pns value), net counts, and photon fluxes (in eight different energy bands) are listed in Tables~\ref{srclist}-\ref{fluxes}. Both \cite{Georgakakis+08} and \cite{Tullmann+11} have found that using a \pns threshold of $4\times10^{-6}$ results in $\sim$0.5 false sources per megapixel. Given the survey size of our NGC~404 observations, we expect there to be $\sim$1.6 false sources included in our final source catalog.

\begin{table*}[ht]
\centering
\caption{The NGC~404 Source List}
\begin{tabular}{cccccccc}
\hline \hline
Source & Source ID & R.A. (J2000) & Decl. (J2000) & Positional   & Total exp. map      & $R_{\rm src}$ & $\theta$  \\
No.        &                    & ($^{\circ}$)    & ($^{\circ}$)      & Error (\asn) & value (s cm$^2$) & (sky pixel)        & (\am)       \\
(1)          & (2)              & (3)                  & (4)                    & (5)                 & (6)                          & (7)                      & (8)            \\
\hline
1        & 010838.25+354027.3 &        17.159396 &        35.674264 &  0.23 &   3.803$\times10^7$ & 21.68 & 10.0 \\
2        & 010841.45+354056.8 &        17.172710 &        35.682458 &  0.47 &   4.848$\times10^7$ & 19.21 &  9.3 \\
3        & 010842.38+354300.8 &        17.176591 &        35.716903 &  0.55 &   4.835$\times10^7$ & 17.38 &  8.9 \\
4        & 010847.72+353948.3 &        17.198868 &        35.663419 &  0.69 &   4.904$\times10^7$ & 15.91 &  8.4 \\
5        & 010848.00+354437.4 &        17.200024 &        35.743725 &  0.37 &   3.429$\times10^7$ &  5.54 &  7.9 \\
\hline\hline
\end{tabular}\label{srclist}
\tablecomments{Column 2: the source ID also contains the source coordinates (J2000.0). Column 5: the positional uncertainty is a simple error circle (in \asn) around the source R.A. and decl. Column 6: sum of mean exposure map values in the source region. Columns 7: the average radius of the source extraction region (1 sky pixel = 0\as392). Column 8: off-axis angle.\newline
Only the first five entries are shown.}
\end{table*}

\begin{table*}[ht]
\centering
\caption{Merged log(\pns) Values for Different Energy Bands}
\begin{tabular}{ccccccccc}
\hline \hline
Source & log(\pns[1])	& log(\pns[2])	& log(\pns[3])	& log(\pns[4])	& log(\pns[5])	& log(\pns[6])	& log(\pns[7])	& log(\pns[8])                 \\
No.        & (0.5-8.0 keV) & (0.5-2.0 keV) & (2.0-8.0 keV) & (0.35-8.0 keV) & (0.35-1.1 keV) & (1.1-2.6 keV) & (2.6-8.0 keV) & (0.35-2.0 keV)    \\
(1)          & (2)                  & (3)                   & (4)                    & (5)                      & (6)                      & (7)                   & (8)                   & (9)                          \\
\hline
1        & $<$-10 & $<$-10 & $<$-10 & $<$-10 & $<$-10 & $<$-10 & $<$-10 & $<$-10 \\
2        & $<$-10 & $<$-10 & $<$-10 & $<$-10 &   -2.79 & $<$-10 & $<$-10 & $<$-10 \\
3        & $<$-10 & $<$-10 &   -4.61 & $<$-10 &   -9.29 & $<$-10 &   -3.78 &   -3.78 \\
4        &   -9.40 & $<$-10 &   -1.40 &   -9.52 &   -5.73 &   -7.64 &   -1.24 &   -1.24 \\
5        & $<$-10 & $<$-10 & $<$-10 & $<$-10 &   -7.96 & $<$-10 &   -6.47 &   -6.47 \\
\hline\hline
\end{tabular}\label{pns}
\tablecomments{The (logarithmic) \pns value is the Poisson probability of not being a source. Log(\pns) values smaller than -10 have been replaced by $<$-10. All these cases are highly significant detections.\newline
Only the first five entries are shown.}
\end{table*}

\begin{table*}[ht]
\centering
\caption{Total Net Counts in Different Energy Bands}
\begin{tabular}{ccccccccc}
\hline \hline
Source & {\it net\_cnts}[1]  & {\it net\_cnts}[2] & {\it net\_cnts}[3] & {\it net\_cnts}[4] & {\it net\_cnts}[5] & {\it net\_cnts}[6] & {\it net\_cnts}[7] & {\it net\_cnts}[8] \\
No.        & (0.5-8.0 keV)     & (0.5-2.0 keV)      & (2.0-8.0 keV)     & (0.35-8.0 keV)  & (0.35-1.1 keV)   & (1.1-2.6 keV)     & (2.6-8.0 keV)     & (0.35-2.0 keV)    \\
(1)          & (2)                       & (3)                        & (4)                       & (5)                       & (6)                        & (7)                       & (8)                       & (9)                          \\
\hline
1        &   648.0$^{+   27.0}_{-   26.0}$ &   442.6$^{+   22.2}_{-   21.2}$ &   205.5$^{+   16.1}_{-   15.0}$ &   655.3$^{+   27.2}_{-   26.1}$ &   166.3$^{+   14.1}_{-   13.0}$ &   341.8$^{+   19.7}_{-   18.7}$ &   147.3$^{+   13.9}_{-   12.8}$ &   147.3$^{+   13.9}_{-   12.8}$ \\
2        &   116.1$^{+   13.0}_{-   12.0}$ &    51.6$^{+    8.7}_{-    7.7}$ &    64.5$^{+   10.3}_{-    9.2}$ &   116.5$^{+   13.1}_{-   12.0}$ &     7.3$^{+    4.4}_{-    3.3}$ &    61.7$^{+    9.4}_{-    8.3}$ &    47.4$^{+    9.1}_{-    8.0}$ &    47.4$^{+    9.1}_{-    8.0}$ \\
3        &    66.9$^{+   10.5}_{-    9.5}$ &    47.1$^{+    8.3}_{-    7.3}$ &    19.8$^{+    7.2}_{-    6.1}$ &    67.4$^{+   10.6}_{-    9.5}$ &    16.1$^{+    5.4}_{-    4.3}$ &    35.1$^{+    7.5}_{-    6.4}$ &    16.2$^{+    6.6}_{-    5.5}$ &    16.2$^{+    6.6}_{-    5.5}$ \\
4        &    32.6$^{+    8.2}_{-    7.1}$ &    25.3$^{+    6.5}_{-    5.4}$ &     7.3$^{+    5.7}_{-    4.6}$ &    33.2$^{+    8.3}_{-    7.2}$ &    10.6$^{+    4.7}_{-    3.6}$ &    16.4$^{+    5.7}_{-    4.5}$ &     6.2$^{+    5.3}_{-    4.2}$ &     6.2$^{+    5.3}_{-    4.2}$ \\
5        &    34.3$^{+    7.1}_{-    6.0}$ &    19.6$^{+    5.6}_{-    4.4}$ &    14.7$^{+    5.1}_{-    4.0}$ &    34.2$^{+    7.1}_{-    6.0}$ &     6.8$^{+    3.8}_{-    2.6}$ &    18.6$^{+    5.4}_{-    4.3}$ &     8.9$^{+    4.3}_{-    3.1}$ &     8.9$^{+    4.3}_{-    3.1}$ \\
\hline\hline
\end{tabular}\label{netcnts}
\tablecomments{Only the first five entries are shown.}
\end{table*}

\begin{table*}[ht]
\centering
\caption{Photon Flux (photons cm$^{-2}$ s$^{-1}$) in Different Energy Bands}
\begin{tabular}{ccccccccc}
\hline \hline
Source & log({\it flux}[1])	& log({\it flux}[2])	& log({\it flux}[3])	& log({\it flux}[4])	& log({\it flux}[5])	& log({\it flux}[6])	& log({\it flux}[7])	& log({\it flux}[8])        \\
No.        & (0.5-8.0 keV) & (0.5-2.0 keV) & (2.0-8.0 keV) & (0.35-8.0 keV) & (0.35-1.1 keV) & (1.1-2.6 keV) & (2.6-8.0 keV) & (0.35-2.0 keV) \\
(1)          & (2)                   & (3)                   & (4)                   & (5)                      & (6)                      & (7)                   & (8)                   & (9)                       \\
\hline
1        & -4.38 & -4.72 & -4.82 & -4.36 & -4.77 & -4.87 & -4.95 & -4.95 \\
2        & -5.23 & -5.76 & -5.43 & -5.22 & -6.23 & -5.72 & -5.55 & -5.55 \\
3        & -5.46 & -5.78 & -5.94 & -5.45 & -5.85 & -5.95 & -6.01 & -6.01 \\
4        & -5.79 & -6.07 & -6.39 & -5.78 & -6.08 & -6.30 & -6.45 & -6.45 \\
5        & -5.26 & -5.66 & -5.58 & -5.25 & -5.66 & -5.74 & -5.78 & -5.78 \\
\hline\hline
\end{tabular}\label{fluxes}
\tablecomments{Only the first five entries are shown.}
\end{table*}
   
\subsection{Sensitivity Maps}
Subsequent analyses in this work require sensitivity maps, which provide the energy flux level at which a source could be detected for each point in our survey area. \AEx provides an estimate of the photon flux for each source ({\it flux2}), which is based on the net source counts, exposure time, and the mean ARF in the given energy band. We convert these photon fluxes to energy fluxes assuming a power law with \PL=1.9 that is absorbed by the Galactic column (\nH=5.13$\times10^{20}$ cm$^{-2}$). Hereafter, we refer to this as the ``standard model.'' Roughly 80\% of our X-ray sources had their spectra automatically fit with \AEx (see Section~\ref{spectra}), and most were consistent with a power law model (\PL$\sim$1.7-1.9) with no absorption beyond the Galactic column, as would be expected for a population dominated by XRBs and background AGN. We do not expect the standard model to systematically bias the \lognlogs distribution towards higher or lower fluxes.

To create sensitivity maps, we calculated the number of source counts that would be required to meet our \pns criteria ($<4\times10^{-6}$) for each point in the survey area. This method for generating sensitivity maps has been tested using simulated \Chandra observations by \cite{Georgakakis+08}. The 90\% complete limiting flux in the 0.35-8 keV, 0.5-2 keV, and 2-8 keV energy bands are 6.2$\times10^{-16}$ \flux, 1.3$\times10^{-16}$ \flux, and 8.7$\times10^{-16}$ \flux, respectively. At the distance of NGC~404, these fluxes correspond to luminosities of 6.3$\times10^{35}$ \lum, 1.6$\times10^{35}$ \lum, and 1.0$\times10^{36}$ lum, respectively. Table~\ref{limitlum} summarizes the limiting fluxes for the 70\%, 90\%, and 95\% completeness levels for our total energy range, the soft band, and the hard band. For comparison, we additionally constructed sensitivity maps using the CIAO task \texttt{lim\_sens}, and found the predicted limiting fluxes to be $\sim$30\% higher than found using our \pns analysis.   

\begin{table*}[ht]
\centering
\caption{Limiting Unabsorbed Fluxes for the Combined NGC~404 Observations$^a$}
\begin{tabular}{cccccc}
\hline \hline
Energy Band & 70\%			& 90\%			& 95\%			\\
(keV)	& ($10^{-16}$ \flux)	& ($10^{-16}$ \flux)	& ($10^{-16}$ \flux) 	\\
(1)		& (2)				& (3)				& (4)				\\
\hline
0.35-8	& 5.0				& 6.2			& 19.8			\\
0.5-2		& 0.3				& 1.3			& 4.4				\\
2-8		& 2.8				& 8.7			& 27.5			\\
\hline\hline
\end{tabular}\label{limitlum}
\tablecomments{$^a$Limiting fluxes are estimated assuming the standard model (\PL=1.9).}
\end{table*}

\section{The X-ray Source Catalog}\label{catalog}
The X-ray properties of individual sources, such as their temporal variability and spectral shape, can be used to constrain the physical origin of the X-ray emission. Additionally, bulk population properties (such as the radial distribution of sources) enable us to identify different X-ray populations in a statistical sense. We expect the X-ray point source population of NGC~404 to consist of LMXBs, background AGN, and foreground stars. To match each source with a likely physical origin, we use our catalog to search for time-variable sources, analyze hardness ratios (for faint sources) and perform spectral fitting (for bright sources), construct a radial source distribution, and create XLFs. The addition of overlapping \HST fields allows us to identify optical counterpart candidates.

\subsection{Time Variable Sources}
The full NGC~404 catalog was systematically searched for sources showing both significant short-term (i.e., variability that occurs on timescales less than the exposure time, $\sim$27 hours) and long-term variability in the 0.35-8 keV band. While rapid X-ray variability is routinely observed in both AGN and XRB systems \citep{Vaughan+03}, it is typically not observed in other sources of X-ray emission (i.e., SNRs). We used a Kolmogorov-Smirnov (K-S) test to compare the cumulative photon arrival time distribution for each X-ray source to a uniform count rate model. The K-S test returns the probability (denoted by $\xi$) that both distributions were drawn from the same parent distribution. We searched for rapid variability only in Obs ID 12239, due to its significantly longer exposure time than the archival observations. We used the three separate \Chandra observations to search for significant changes in the long-term X-ray flux of each source.

Only two sources, 1 and 72, have K-S probabilities of 0.06\% and 0.09\% respectively, which may be indicative of potential short-term variability. To investigate the long-term variability of each of the NGC~404 sources, we utilize the variability threshold $\eta$ \citep[as in ][]{Tullmann+11}, defined as $\eta=(flux_{\rm max}-flux_{\rm min})/\Delta flux$, where $flux_{\rm max}$ and $flux_{\rm min}$ are the maximum and minimum observed fluxes of the source, respectively, and the flux error $\Delta flux$ is calculated using the Gehrels approximation \citep[appropriate for low count data; ][]{Gehrels86}. We consider any sources with $\eta\ge3$ potentially variable. One source (\#16) in NGC~404 has $\eta\ge5$ and is highly likely to exhibit long-term variability.

We also consider the possibility that the NGC~404 catalog may contain transient sources. We found 18 X-ray sources that were detected in at least one exposure, but were not detected in another exposure (despite the source's position being within the field of view). We use these non-detections to place an upper limit on the source flux at the time of the exposure; some of these sources are faint, and so may have been just below the detection limit of the shallower exposure without requiring a significant change in flux. However, seven X-ray sources exhibited a change in X-ray flux of at least an order of magnitude and had $\eta\ge3$ -- we refer to these sources as candidate X-ray transients (XRTs). Both LMXBs and HMXBs are known to experience X-ray outbursts of this magnitude \citep{Bozzo+08,Fragos+08,Paul+11,Revn+11}, and may have implications for the slope of the XLF (see \S~\ref{sXLF}).

Table~\ref{variable} lists the variability properties for all sources in the NGC~404 catalog that exhibit some form of temporal variability. We include three potentially variable sources (sources 27, 49, and 57, noted with italics) which fall short of our definition of an XRT candidate by less than 10\%. That is, if any of these three sources were to change either their lowest or highest observed fluxes by less than 10\%, the fractional change in flux would be more than an order of magnitude (the typical flux error for an NGC~404 source is $\sim$20-25\%). 

\begin{table*}[ht]
\centering
\caption{Temporal Variability of NGC~404 X-ray Sources}
\begin{tabular}{cccc}
\hline \hline
Source No.	& $\xi^a$	& $\eta^b$	& $flux_{max}/flux_{min}^c$	\\
(1)          		& (2)					& (3)		& (4)			\\
\hline
1 	& {\bf $<$0.01} 	& 0.10	& 2.61	\\
16 	& 0.09 	& {\bf 5.21}	& {\bf 25.28}	\\
18 	& 0.87 	& 2.03	& {\bf 14.74}	\\
27 	& 0.94 	& 1.81	& {\it 9.21}	\\ 
37 	& 0.25 	& 2.61	& {\bf 10.12}	\\
40 	& 0.04 	& {\bf 3.44}	& {\bf 14.74}	\\
49 	& 0.29 	& 2.10	& {\it 9.46}	\\
53 	& 0.92 	& 1.90	& {\bf 13.78}	\\
54 	& 0.69 	& {\bf 3.02}	& {\bf 23.53}	\\
55 	& 0.79 	& {\bf 4.22}	& {\bf 42.64}	\\
57 	& 0.56 	& 2.00	& {\it 9.19}	\\
59 	& 0.49 	& {\bf 4.96}	& {\bf 23.17}	\\
61 	& 0.89 	& {\bf 3.37}	& {\bf 26.60}	\\
64 	& 1.00 	& {\bf 4.65}	& {\bf 42.46}	\\
71	& 0.06 	& {\bf 4.46}	& {\bf 34.79}	\\
72 	& {\bf $<$0.01} 	& 0.11	& 1.13	\\
\hline\hline
\end{tabular}\label{variable}
\tablecomments{\newline
$^a$The K-S probability of the source being constant within our \Chandra observation Obs ID 12239 (short-term variability). Sources with K-S probabilities of $\xi\leq3.7\times10^{-3}$ are likely to show short-term variability and are shown in boldface.\newline
$^b$$\eta$ is the variability index defined by \cite{Tullmann+11} and is sensitive to long-term variability. Sources with $\eta\ge3$ are likely to show variability between the individual \Chandra exposures.\newline
$^c$Sources whose fluxes changed by more than an order of magnitude are shown in boldface; sources just short of this threshold (by less than 10\%) are shown in italics.}
\end{table*}

\subsection{Radial Source Distribution}\label{radial_src}
While direct, multiwavelength detection of X-ray source counterparts provides an excellent means of discriminating between sources associated with NGC~404 and background or foreground contaminants, this approach is likely to fail when classifying faint sources. As a result, the physical nature of many faint X-ray sources in our catalog will remain ambiguous. We therefore adopt an additional, statistical approach to constrain the number of sources in NGC~404.

\begin{figure*}
\centering
\begin{tabular}{cc}
\includegraphics[width=0.5\linewidth,clip=true,trim=2cm 12.5cm 2cm 2.5cm]{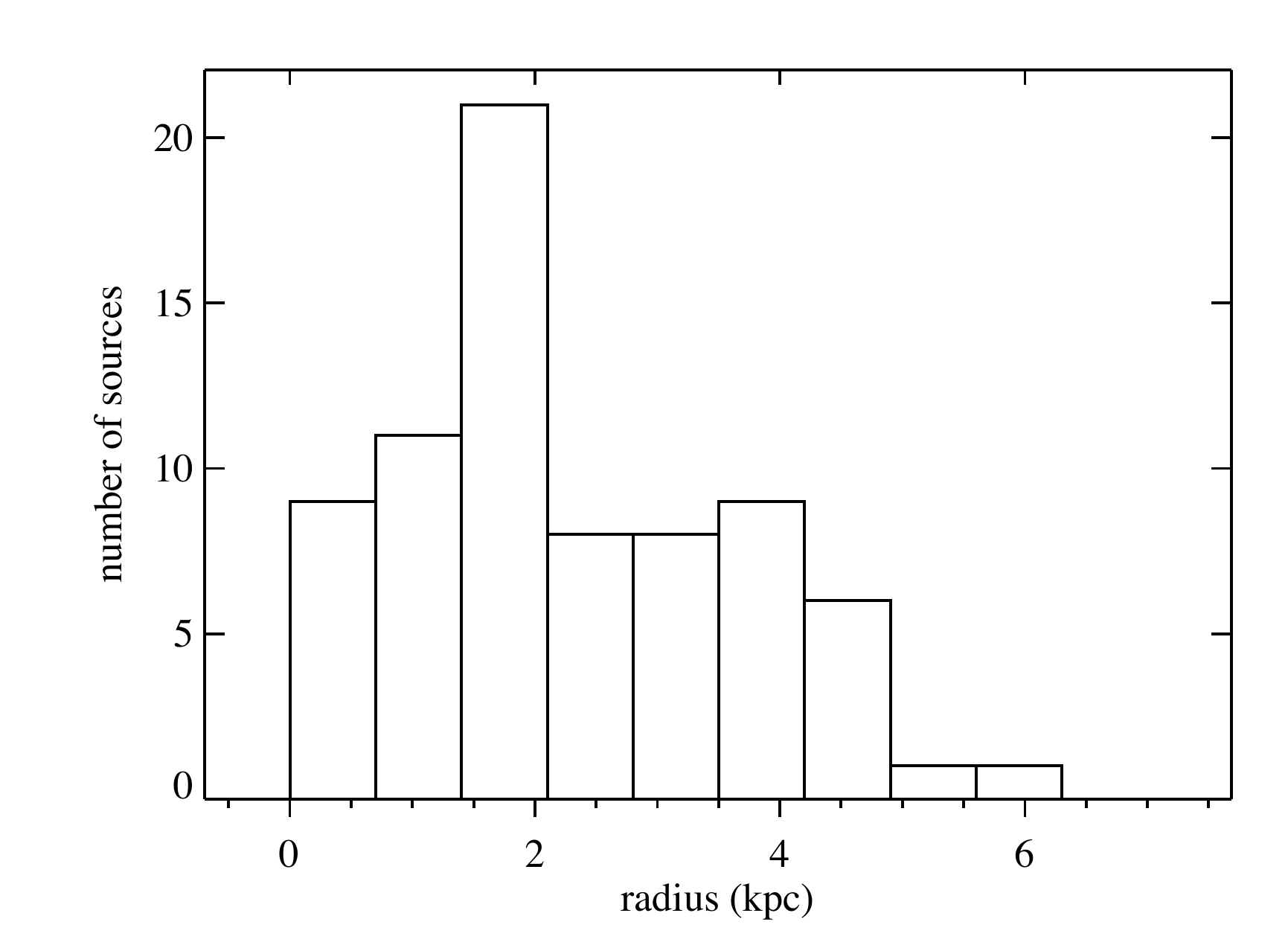} & 
\includegraphics[width=0.5\linewidth,clip=true,trim=2cm 12.5cm 2cm 2.5cm]{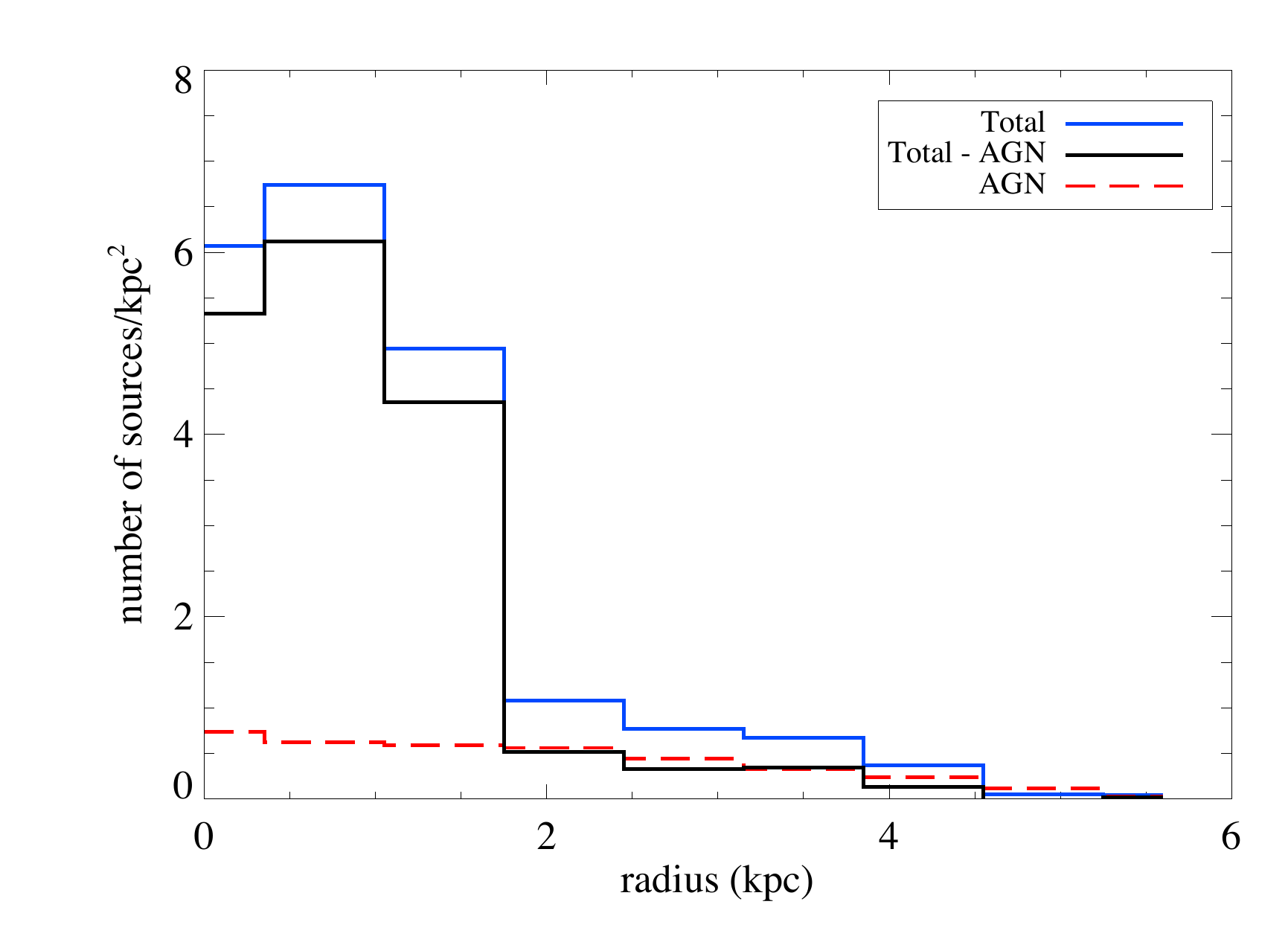} \\
\end{tabular}
\caption{Left: histogram of galactocentric distances for the NGC~404 X-ray source catalog. Right: galactocentric source density profile of NGC~404 in the 0.35-8.0 keV band. The blue histogram shows the total number of sources per square kiloparsec. The red dashed line shows the anticipated number of background AGNs (corrected for the \Chandra ACIS off-axis sensitivity), while the black histogram shows the expected number of sources after correcting for background AGNs.}
\label{radialdist}
\end{figure*}

We construct an X-ray galactocentric profile of the NGC~404 by summing number of sources detected as a function of galactocentric radius. We assign an inclination-corrected galactocentric distance to each X-ray sources assuming the galaxy center is located at $\alpha_0$ = 17\mbox{\ensuremath{.\!\!^{\circ}}}362500, $\delta_0$ = +35\mbox{\ensuremath{.\!\!^{\circ}}}718056 \citep{Evans+10}, and $i=11^{\circ}$ \citep{delRio+04}. The maximum radius we find for an X-ray source in our \Chandra field is $r=6.1$ kpc. We divide the X-ray sources into radial bins (spaced 0.7 kpc apart) based on their distance from the center of the galaxy. We use the cumulative \lognlogs distribution of \cite{Cappelluti+09}, corrected for the decreasing \Chandra sensitivity with increasing off-axis angle, to estimate the expected contamination by background AGN. This was done by finding the number of AGN predicted (for the flux sensitivity of our survey) for the area spanned by each radial bin. In Figure~\ref{radialdist}, we show both a histogram of our derived galactocentric distances and the number of X-ray sources per square kpc. 

The majority of the NGC~404 X-ray sources have $r\lesssim2.5$ kpc, and the number of X-ray sources per kpc$^2$ beyond 2.5 kpc is roughly equal to the number predicted by the AGN \lognlogs distribution. We therefore expect the 29 X-ray sources with $r>2.5$ kpc to be predominantly contaminating background AGN or foreground stars, while the X-ray sources with $r<2.5$ kpc are expected to be a mix of contaminants and X-ray sources associated with NGC~404. This interpretation is consistent with the observed bulge effective radius ($r_e=0.9\pm0.4$ kpc) and disk scale length ($\sim1.9$ kpc) derived by both \cite{Baggett+98} and \cite{Williams+10}.

\subsection{Hardness Ratios}
\begin{figure*}
\centering
\begin{tabular}{cc}
\includegraphics[width=0.5\linewidth,clip=true,trim=2cm 12.5cm 2cm 2.5cm]{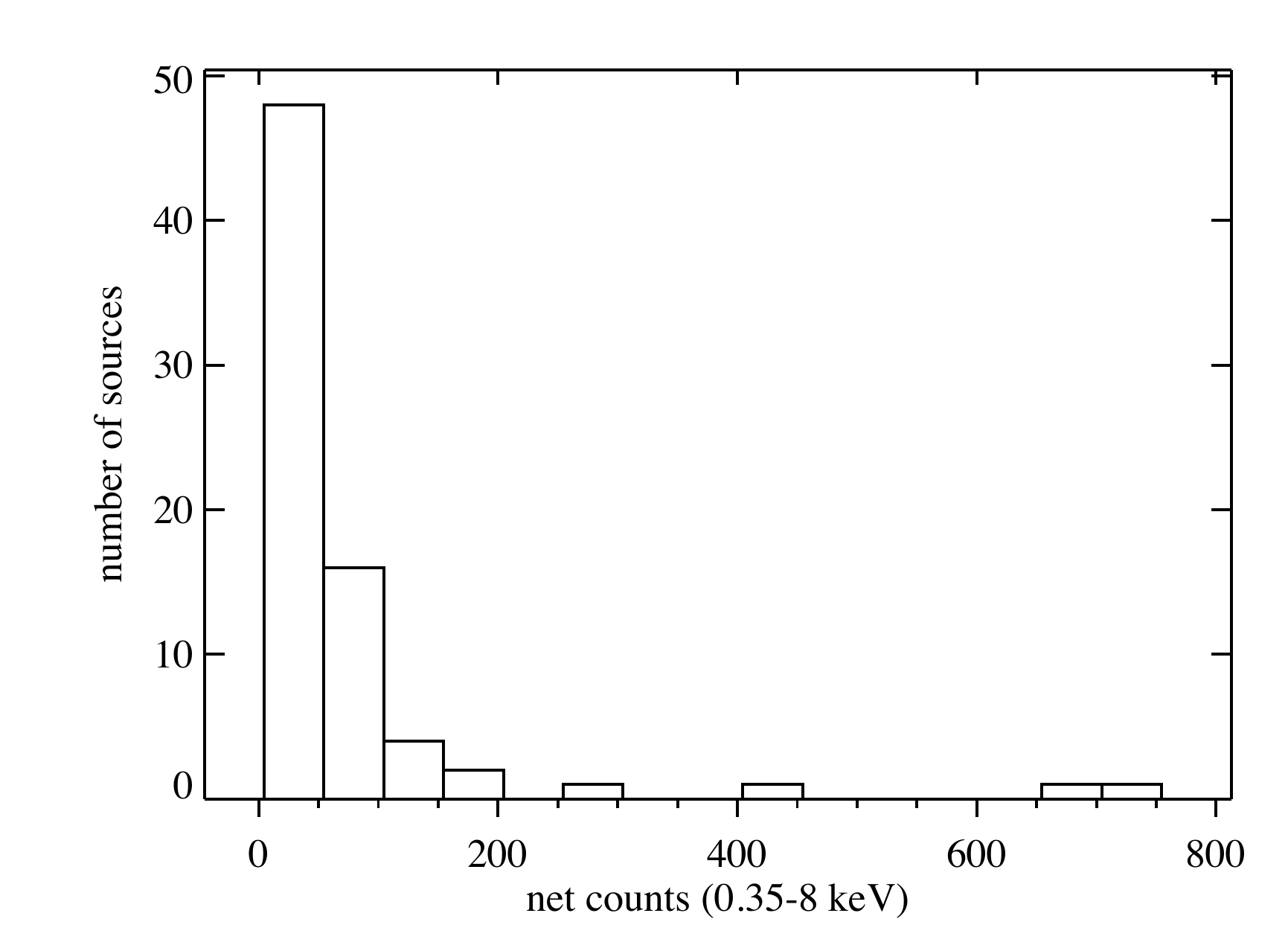} & 
\end{tabular}
\caption{Histogram showing the NGC~404 X-ray source counts. 85\% of X-ray sources in our catalog have fewer than 100 net counts.}
\label{counts_hist}
\end{figure*}

The X-ray spectrum of a source can provide a key diagnostic for separating different populations. For example, LMXBs and AGN are well-described by power laws with photon indices ranging from $\Gamma\sim1-2$, and may show high levels of intrinsic absorption (i.e., in the case of obscured AGN). Super-soft sources (SSSs) and contaminating foreground stars will exhibit significantly softer, thermal X-ray emission with plasma temperatures of a few keV or less. However, when studying XRBs at distances of a few Mpc it is near impossible to constrain spectral parameters to any degree of accuracy, especially for those sources with $\lesssim50$ counts. The majority of NGC~404 X-ray sources fall within this low-count regime, as shown in Figure~\ref{counts_hist}.

Instead of directly measuring the X-ray spectral shape, we define hardness ratios (HRs; also called X-ray colors) to separate different populations. HRs measure the fraction of photons emitted by a source in three energy ranges: soft ($S$, 0.35-1.1 keV), medium ($M$, 1.1-2.6 keV), and hard ($H$, 2.6-8 keV). We evaluate two HRs for each source using the approach developed in \cite{Tullmann+11} \citep[see also][]{Prestwich+09}. Source and background counts were determined by \AEx in each band, and we define a ``soft'' color,

\begin{equation}
HR1 = \frac{M-S}{H+M+S},
\end{equation}

\noindent and a ``hard'' color,

\begin{equation}
HR2 = \frac{H-M}{H+M+S}.
\end{equation}

We use the Bayesian Estimation of Hardness Ratios\footnote{\url{http://hea-www.harvard.edu/AstroStat/BEHR/}} \citep[{\it BEHR}, ][]{Park+06} in a similar manner as described by \cite{Tullmann+11} and \cite{Prestwich+09} to determine the hardness ratios of the NGC~404 X-ray sources. For low-count sources (which make up the majority of the NGC~404 point source catalog), the net counts can be negative due to fluctuations in the source and background estimates, resulting in unphysical HRs when calculated using a traditional approach. The {\it BEHR} code accounts for the fact that source and background counts are non-negative, ensuring that our approach produces physically allowable values of the HRs. In the high-count regime, the results of the Bayesian approach become equivalent to traditionally-computed HRs.

For each source, the inputs are source counts, background counts, the \AEx ``backscale'' parameter (which accounts for the ratio of the source and background extraction areas and efficiencies), and a factor converting from counts to photon flux (i.e., the exposure time multiplied by the mean ARF over the extraction region in the respective energy band). The ``softeff'' and ``hardeff'' parameters take into account the variations in effective area and exposure times between the different observations, especially important considering the strong differences in ACIS filter contamination between the observations utilized in this work; we set these parameters to the exposure time multiplied by the mean ARF values computed by \AEx in the appropriate energy band for each observation in which the source was observed. 
    
The {\it BEHR} code is then run twice (once for $S$ and $M$, and again for $S$ and $H$). For each energy band, we set the {\it BEHR} ``burnin'' parameter to 50,000 and the ``total draws'' parameter to 100,000. This provides 50,000 samples from the probability distribution for each energy band. Combining the results from the two {\it BEHR} runs, we obtain 50,000 samples from the probability distribution for the $S$, $M$, and $H$ counts. Using these distributions, we compute 50,000 values for $HR1$ and $HR2$ for each source. Because the difference of two Poisson distributions does not follow Poisson statistics, we do not use the error estimates directly provided by these {\it BEHR} runs. Instead, the HR value is defined as the mean of the distribution, and the credible interval is evaluated based on the 68.2\% equal-tail estimates (i.e., 0.682/2 of the samples have values below the lower limit, and 0.682/2 of the samples have values above the upper limit). 

\begin{figure*}
\centering
\includegraphics[width=1\linewidth,angle=0,clip=true,trim=2cm 12.5cm 2cm 2cm]{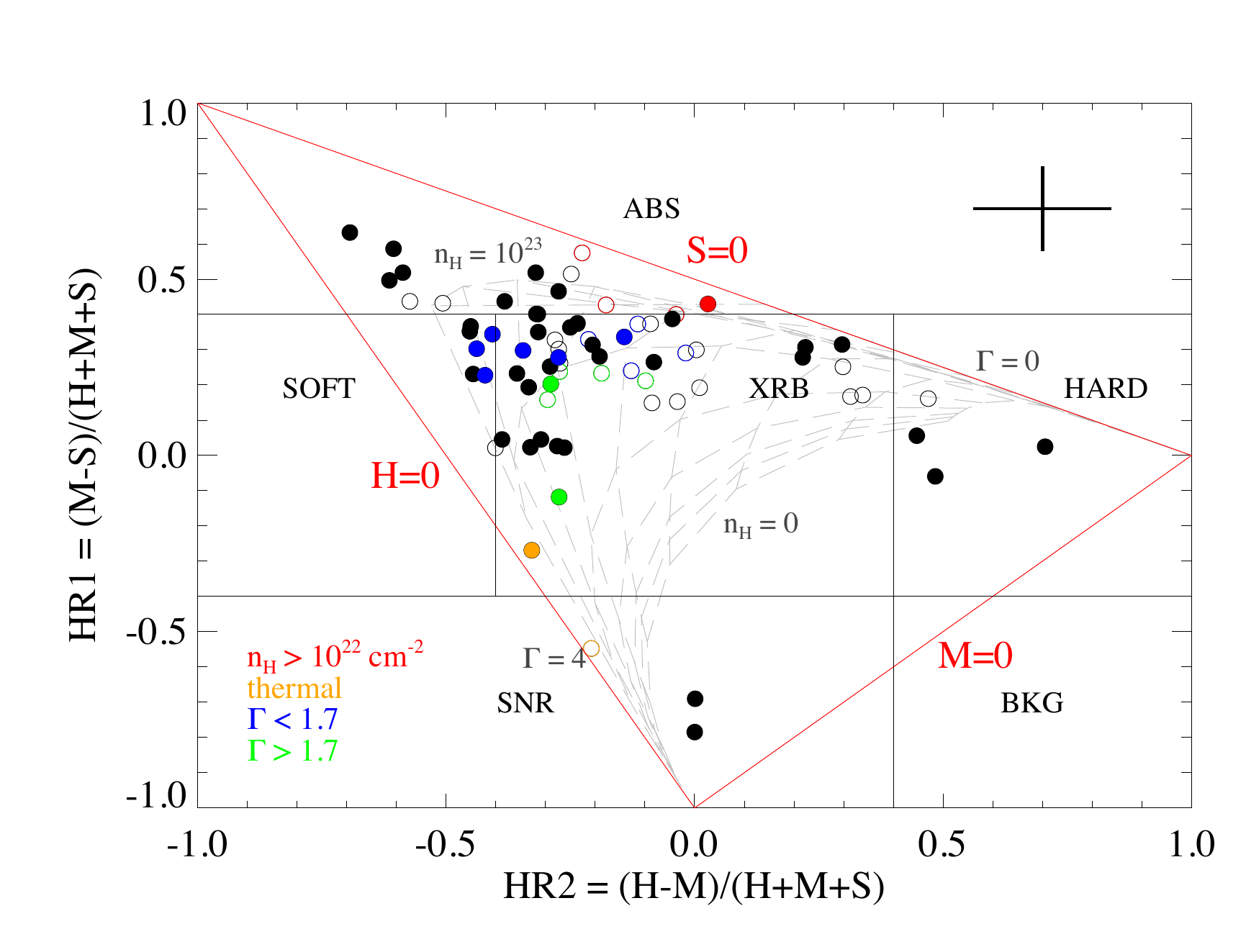} 
\caption{The X-ray color-color diagram. The cross in the upper-right shows the typical error size. Preliminary source classification regions are labeled. Red lines indicate the zero-count limits in the hard (`H'), medium-hard (`M'), and soft (`S') bands -- HRs exterior to this red triangle are non-physical. Filled circles show sources with galactocentric radii less than $\sim2.5$ kpc, while open circles show sources beyond 2.5 kpc. Sources are color-coded according to their best-fit spectral model (see Section~\ref{spectra}): red indicates heavy absorption, orange shows thermal sources, blue shows sources modeled as power laws with $\Gamma<1.7$, and green shows sources modeled as power laws with $\Gamma>1.7$. The gray lines show the X-ray colors expected for a variety of power law models (ranging from \PL=0-4 in steps of 0.5) with different absorbing columns (ranging from \nH = 0 to 10$^{23}$ cm$^{-2}$, in steps of 2.5$\times10^{20}$ cm$^{-2}$).}
\label{HR}
\end{figure*}

To aid in the interpretation of our HR calculations, we use the same X-ray color-color classification scheme developed in Paper I. We define six categories of X-ray sources: `XRB' (which is likely contaminated with background AGN, given their X-ray similarities to LMXBs), `BKG' for likely background sources with an indeterminate spectral shape, `ABS' for heavily absorbed sources, `SNR' (although, given the morphology and SFH of NGC~404, these sources are likely to be foreground stars or super-soft sources, and not true supernova remnants), and indeterminate `HARD' and `SOFT' sources. Our X-ray color-color classification scheme, along with the number of sources belonging to each category, is given in Table~\ref{color_class}. The HRs and corresponding source classification is given in Table~\ref{HRtable}, and a plot of our HR values is shown in Figure~\ref{HR}. Figure~\ref{HR} additionally shows the agreement between the HRs and spectral fitting (see Section~\ref{spectra}) -- HRs provide a useful technique for identifying highly absorbed sources, sources with thermal emission, and sources with power law-like spectra. However, HRs cannot be used to separate power laws with different photon indices.
  
We find $\sim$61\% of the NGC~404 X-ray sources fall within the `XRB'-like category, and $\sim$20\% of sources show evidence for high levels of intrinsic absorption. Only four sources (5\%) are designated as `SNR'-like -- one of these sources (source 65) is coincident with a foreground star (see Section~\ref{spectra}). Given the latitude of NGC~404 and the small area subtended by the galaxy, these are likely foreground objects not associated with NGC~404. The remaining X-ray sources are categorized as `HARD' or `SOFT' and are likely not associated with NGC~404.

\begin{table*}[ht]
\centering
\caption{Hardness Ratio Source Classification Scheme}
\begin{tabular}{ccc}
\hline \hline
Classification & Definition & \# Sources \\
(1) & (2) & (3) \\
\hline
X-ray binary (`XRB')				& $-0.4 < HR2 < 0.4$, $-0.4 < HR1 < 0.4$	& 45 \\
Background source (`BKG')		& $HR2 > -0.4$, $HR1 < -0.4$			& 0 \\
Absorbed source (`ABS')			& $HR1 > 0.4$						& 15 \\
Supernova remnant (`SNR')		& $HR2 < -0.4$, $HR1 < -0.4$			& 3 \\
Indeterminate hard source (`HARD')	& $HR2 > 0.4$, $-0.4 < HR1 < 0.4$		& 4 \\
Indeterminate soft source (`SOFT')	& $HR2 < -0.4$, $-0.4 < HR1 < 0.4$		& 7 \\
\hline \hline
\label{color_class}
\end{tabular}
\end{table*}

\begin{table*}[ht]
\centering
\caption{Hardness Ratios}
\begin{tabular}{cccc}
\hline \hline
Source No. & HR1 & HR2 & Classification \\
(1) & (2) & (3) & (4) \\
\hline
1 &  0.26 $\pm$  0.03 & -0.27 $\pm$  0.03 & XRB \\
2 &  0.40 $\pm$  0.05 & -0.04 $\pm$  0.08 & XRB \\
3 &  0.24 $\pm$  0.08 & -0.13 $\pm$  0.09 & XRB \\
4 &  0.15 $\pm$  0.11 & -0.09 $\pm$  0.12 & XRB \\
5 &  0.33 $\pm$  0.13 & -0.28 $\pm$  0.15 & XRB \\
\hline \hline
\label{HRtable}
\end{tabular}
\tablecomments{Only the first five entries are shown.}
\end{table*}

\subsection{X-ray Spectral Analysis}\label{spectra}
For sufficiently bright sources ($\gtrsim50$ counts), the X-ray spectrum of a source can be used to constrain the physical nature of the X-ray emission. Both XRBs and AGN are typically described by power laws, with photon indices ranging from \PL$\sim1-2$. If the source is a likely XRB, the photon index can sometimes be used to constrain the nature of the compact object: NS primaries typically exhibit harder X-ray emission, with \PL$\sim$1-1.5, while BHs (which lack a solid surface) often produce \PL$\sim2-2.5$. Foreground stars will show significantly softer X-ray emission, and typically have their spectra fit with a thermal plasma.

Spectra were extracted by \AEx using the \texttt{CIAO} tool \texttt{dmextract}. Response products were created by \AEx using the \texttt{CIAO} tools \texttt{mkacisrmf} and \texttt{mkarf}. Due to the low number of counts found for most of the sources in our catalogs, we fit spectral models to the unbinned spectra and use $C$-statistics in lieu of traditional \chisq statistics \citep[see e.g., ][]{Cash79,Humphrey+09}. Three sources were found have more than 200 net counts in the 0.35-8 keV band; these sources are discussed in more detail below. Detailed spectral fitting was performed for the NGC~404 central engine in \cite{Binder+11a}.

The ungrouped spectra for 60 out of 74 sources in NGC~404 were automatically fit in \AEx with one of four models: a power-law model, an \texttt{APEC} model \citep{Smith+01}, or either a power-law or thermal model with additional absorption. The Galactic absorption was modeled with the \texttt{tbabs} model \citep{Wilms+00} by freezing the Galactic \nH at 5.13$\times10^{20}$ cm$^{-2}$ \citep{Kalberla+05}. In order to determine the best-fit model, models that predicted unreasonably high photon indices (\PL$>4$) or plasma temperatures ($kT>6.0$ keV) were rejected. 

Because the $C$-statistic cannot be directly interpreted as a goodness-of-fit, we use a different approach to select the preferred spectral model for each source. First, we used \texttt{XSPEC}'s \texttt{goodness} command to perform 5,000 Monte Carlo simulations for each spectral model (power law and thermal models, both with and without additional absorption). The \texttt{goodness} command simulates spectra based on the model parameters and finds the number of simulations with a fit statistic less than that found for the data. If the observed spectrum was produced by that model, the ``goodness'' should be $\sim$50\%. For some sources, the choice of preferred spectral model was obvious: three of the spectral models yielded a \texttt{goodness} of less than $\sim2$\%, while one model produced a \texttt{goodness} in the range of $\sim40-60$\%. In the remaining cases, no obviously preferred model was found -- for these sources, we chose the simple, unabsorbed power law unless the spectrum showed indications for emission features (in which case, the unabsorbed thermal model was selected). The best-fit model and parameters for sources with greater than 50 net counts are listed in Table~\ref{spectralfits}. All errors represent the 90\% uncertainties.

\begin{table*}[ht]
\centering
\caption{Best-Fit Spectral Models for Sources with More Than 50 Counts}
\begin{tabular}{ccccccccccc}
\hline \hline
No. & Best-fit$^a$  & \nH$^b$ &	\PL &	$kT$ & 	Norm$^c$ & $dof^d$ & $C$ & $f_X$ (0.5-2.0 keV)$^e$ & $f_X$ (0.5-8.0 keV)$^f$	& \texttt{goodness} \\
       & Model            & ($10^{21}$ cm$^{-2}$) &       & 	(keV) &			&                  &                     & \multicolumn{2}{c}{(10$^{-15}$ \flux)}        		&        \\
 (1) & (2)		     & (3)  & (4)		&  (5)		& (6)		& (7)			& (8)			& (9) & (10)							& (11)	\\
\hline
1 & pow &  $<1.1$  &  1.65$^{+0.12}_{-0.11}$ &  ...  &  2.02$\times10^{-5}$ & 510 & 496 						& 41.1$^{+1.5}_{- 1.4}$ 	& 73.4$^{+ 5.2}_{- 4.1}$ &	 34\%	\\ 
2 & pow & $<12$ &  0.6$\pm0.3$ &  ...  &  1.48$\times10^{-6}$ & 1015 &  943.22 							&  3.8$^{+ 0.7}_{- 2.9}$ 	& 31.1$^{+ 6.2}_{- 13.6}$ &  88\%	\\
3 & pow &  $<1.1$  &  1.5$\pm0.4$ &  ...  &  1.72$\times10^{-6}$ & 1016 &  818.53 							&  1.6$^{+ 0.5}_{- 0.4}$ 	&  4.8$^{+ 1.9}_{- 1.3}$ &	0.4\%	\\ 
7 & pow &  $<1.4$  &  1.8$\pm0.1$ &  ...  &  1.18$\times10^{-5}$ & 510 &  414 								& 65.6$^{+7.3}_{-8.3}$ 	& 40.7$^{+4.6}_{- 4.0}$ & 58\%		\\
11 & pow &  $<19$  &  0.8$\pm0.4$ &  ...  &  8.14$\times10^{-7}$ & 1016 &  964.18 							& 0.9$^{+ 0.4}_{- 0.3}$ 	&  5.5$^{+ 3.9}_{- 1.7}$ &	12\%	\\ 
13 & pow &  $<1.9$  &  1.2$\pm0.3$ &  ...  &  1.83$\times10^{-6}$ & 1016 &  953.24 							& 1.8$\pm0.4$ 			&  7.3$^{+ 1.7}_{- 1.8}$ &	25\%	\\ 
{\bf 17} & pow &  ...  &  1.9$\pm0.2$ &  ...  &  5.17$\times10^{-6}$ & 1016 &  795.63 							&  5.0$^{+ 0.7}_{- 0.6}$ 	& 10.8$^{+ 2.0}_{- 1.5}$ & 60\%	\\
{\bf 20} & pow &  ...  &  1.5$\pm0.3$ &  ...  &  1.70$\times10^{-6}$ & 1016 &  671.21 							&  1.6$^{+ 0.2}_{- 0.4}$ 	&  5.0$^{+ 1.8}_{- 1.0}$ &	2\%	\\ 
{\bf 32} & apec &  ...  &  ... &  0.8$\pm0.1$  &  4.56$\times10^{-6}$ & 521 &  128.13 							&  8.6$^{+ 0.3}_{- 0.5}$ 	&  9.0$^{+ 0.2}_{- 0.3}$ &	63\%	\\ 
{\bf 35} & apec & 37$^{+13}_{-11}$ &  ...  &  2.8$^{+ 3.2}_{- 1.0}$ &  4.77$\times10^{-5}$ & 1015 &  829.52 	& 12.2$^{+ 2.2}_{- 3.6}$ 	& 19.4$^{+ 2.6}_{-10.1}$ & 95\%	\\ 
{\bf 39} & pow &  $<1.3$  &  1.4$\pm0.3$ &  ...  &  1.53$\times10^{-6}$ & 1016 &  717.66 							&  1.5$\pm0.2$ 		&  4.8$^{+ 1.4}_{- 1.1}$ &	66\%	\\
{\bf 41$^g$} & pow &  ...  &  2.1$\pm0.2$ &  ...  &  4.56$\times10^{-6}$ & 1016 &  657.36 						&  4.3$^{+ 0.6}_{- 0.3}$ 	&  8.0$^{+ 1.1}_{- 1.0}$ &	82\%	\\ 
{\bf 45} & pow &  $<1.1$  &  1.6$\pm0.4$ &  ...  &  9.01$\times10^{-7}$ & 1016 &  542.89 							&  0.9$^{+ 0.3}_{- 0.2}$ 	&  2.4$^{+ 0.9}_{- 0.6}$ &	0.1\%	\\
{\bf 46} & pow &  $<2.1$  &  1.4$\pm0.2$ &  ...  &  2.73$\times10^{-6}$ & 1016 &  823.33 							&  2.7$^{+ 0.4}_{- 0.5}$ 	&  8.7$^{+ 1.8}_{- 1.5}$ &	33\%	\\
{\bf 55} & pow &  $<3.1$  &  0.9$\pm0.3$ &  ...  &  8.94$\times10^{-7}$ & 1016 &  845.80 							&  0.9$\pm0.3$ 		&  5.5$^{+ 1.5}_{- 1.3}$ &	62\%	\\ 
58 & pow &  $<5.5$  &  1.5$^{+ 0.4}_{- 0.3}$ &  ...  &  1.10$\times10^{-6}$ & 1016 &  762.39 					&  1.1$\pm0.3$ 		&  3.3$^{+ 1.5}_{- 0.8}$ &	29\%	\\ 
64 & pow &  ...  &  1.8$\pm0.4$ &  ...  &  1.36$\times10^{-6}$ & 1016 &  781.63 							&  1.3$^{+ 0.4}_{- 0.2}$ 	&  3.0$^{+ 1.2}_{- 0.7}$ &	55\%	\\
65 & apec &  ...  &  ...  &  0.57$^{+0.03}_{-0.04}$ &  4.71$\times10^{-5}$ & 98 &  106 					& 40.5$^{+ 2.3}_{- 2.4}$ 	& 41.2$^{+1.5}_{-1.9}$  &	32\%	\\ 
67 & pow &  ...  &  1.2$\pm0.5$ &  ...  &  1.11$\times10^{-6}$ & 1016 &  923.92 							&  1.1$\pm0.3$ 		&  4.7$^{+ 2.0}_{- 1.4}$ &	49\%	\\
68 & pow &  $<4.1$  &  1.2$\pm0.4$ &  ...  &  1.54$\times10^{-6}$ & 1016 &  992.32 							&  1.5$\pm0.4$ 		&  6.4$^{+ 3.2}_{- 1.7}$ &	34\%	\\
70 & pow &  ...  &  1.8$\pm0.5$ &  ...  &  2.00$\times10^{-6}$ & 1016 &  899.26 							&  1.9$\pm0.4$ 		&  4.3$^{+ 1.4}_{- 0.8}$ &	41\%	\\ 
72 & pow &  ...  &  1.7$\pm0.4$ &  ...  &  2.31$\times10^{-6}$ & 1016 & 1071.92 						&  2.3$^{+ 0.6}_{- 0.5}$ 	&  5.8$^{+ 3.2}_{- 1.2}$ &	54\%	\\
73 & pow &  $<9.9$  &  1.1$\pm0.3$ &  ...  &  2.90$\times10^{-6}$ & 1016 & 1108.80 						&  3.0$^{+ 0.7}_{- 0.5}$ 	& 13.1$^{+ 3.3}_{- 2.7}$ &	 72\%	 \\ 
\hline \hline
\label{spectralfits}
\end{tabular}
\tablecomments{Sources with galactocentric radii less than 2.5 kpc are shown in boldface, as these sources are more likely to be associated with NGC~404.\newline
$^a$ The best-fit model is either a \texttt{powerlaw} or \texttt{vapec} model, depending on which one has the lower \Cdof. \newline
$^b$ Intrinsic source absorption, if beyond the Galactic column was required.\newline
$^c$ Normalization constant of the fit, in units of photons keV$^{-1}$ cm$^{-2}$ s$^{-1}$ and cm$^{-5}$. \newline
$^d$ Degrees of freedom. \newline
$^{e,f}$ Unabsorbed X-ray fluxes. \newline
$^g$ The NGC~404 central engine; see \cite{Binder+11a}.}
\end{table*}

The spectra of bright sources often require multiple component models. Three X-ray sources (1, 7, and 65, excluding the NGC~404 nucleus) have more than 450 net counts, however none of these sources is likely to be associated with NGC~404. Sources 1 and 7 are likely background AGN whose spectra do not require more complex models than what is presented in Table~\ref{spectralfits}. 

\subsubsection{Source 65}
Although source 65 (a likely foreground star) does not require a more complex spectral model to adequately describe its X-ray spectrum, it is interesting enough to warrant a brief discussion. Source 65 was detected in two of the observations considered here (Obs IDs 12239 and 384). The X-ray source was detected on the back-illuminated S3 chip in Obs ID 12239, while it was detected on the front-illuminated S4 detector in Obs ID 384. The differences in front and back-illuminated responses make merging the source 65 spectrum problematic; we therefore only the spectrum extracted from the 97 ks Obs ID 12239 observation (which is much higher quality than the 1.8 ks snapshot taken in Obs ID 384). Source 65 has a large off-axis angle from the nominal aimpoint of Obs ID 12239, $\sim$6\am8.

\begin{figure*}
\centering
\begin{tabular}{cc}
\includegraphics[width=0.37\linewidth,angle=0,clip=true,trim=0cm 0cm 0cm 0cm]{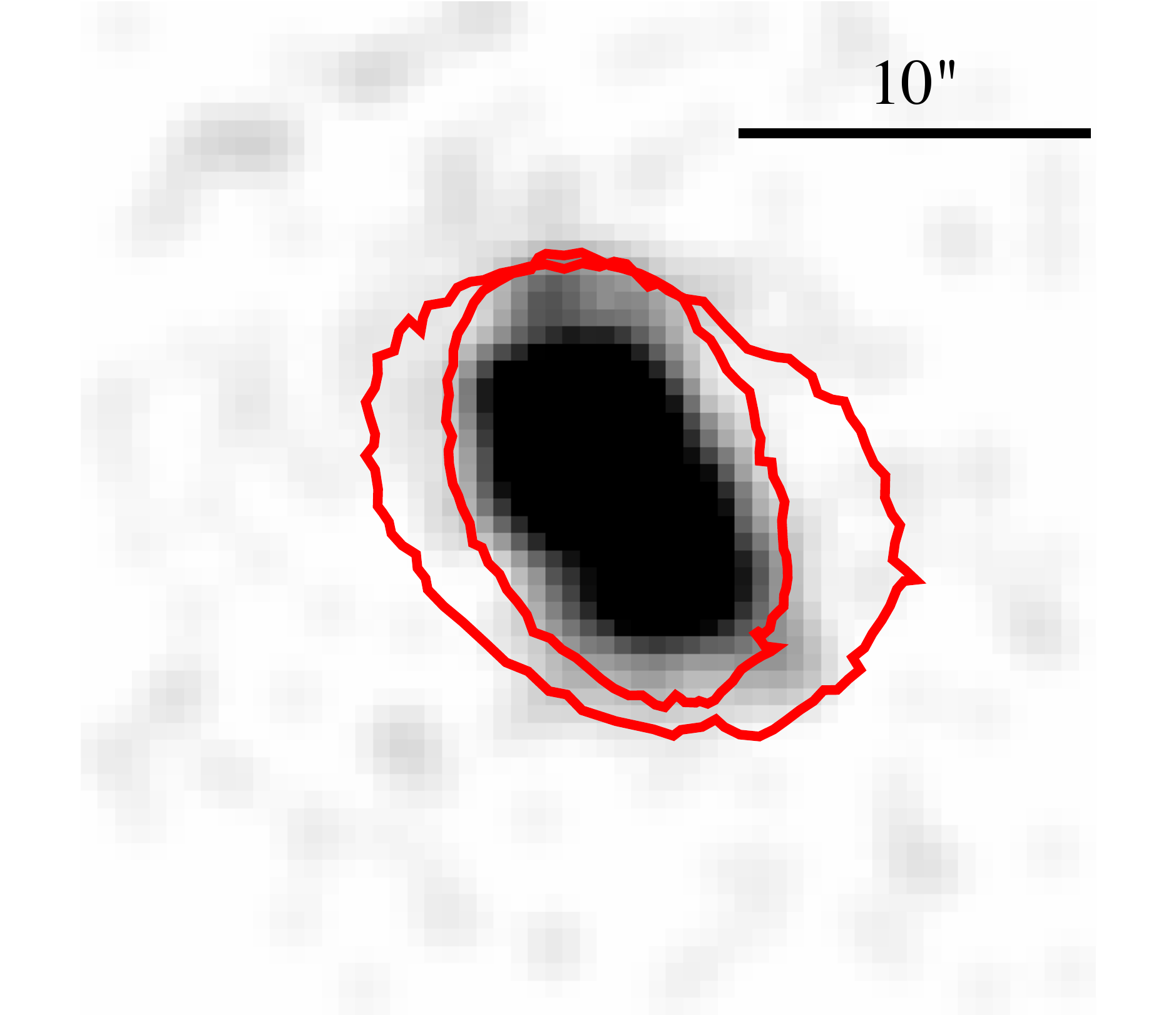} &
\includegraphics[width=0.45\linewidth,angle=0,clip=true,trim=1.1cm 0cm 0cm 0cm]{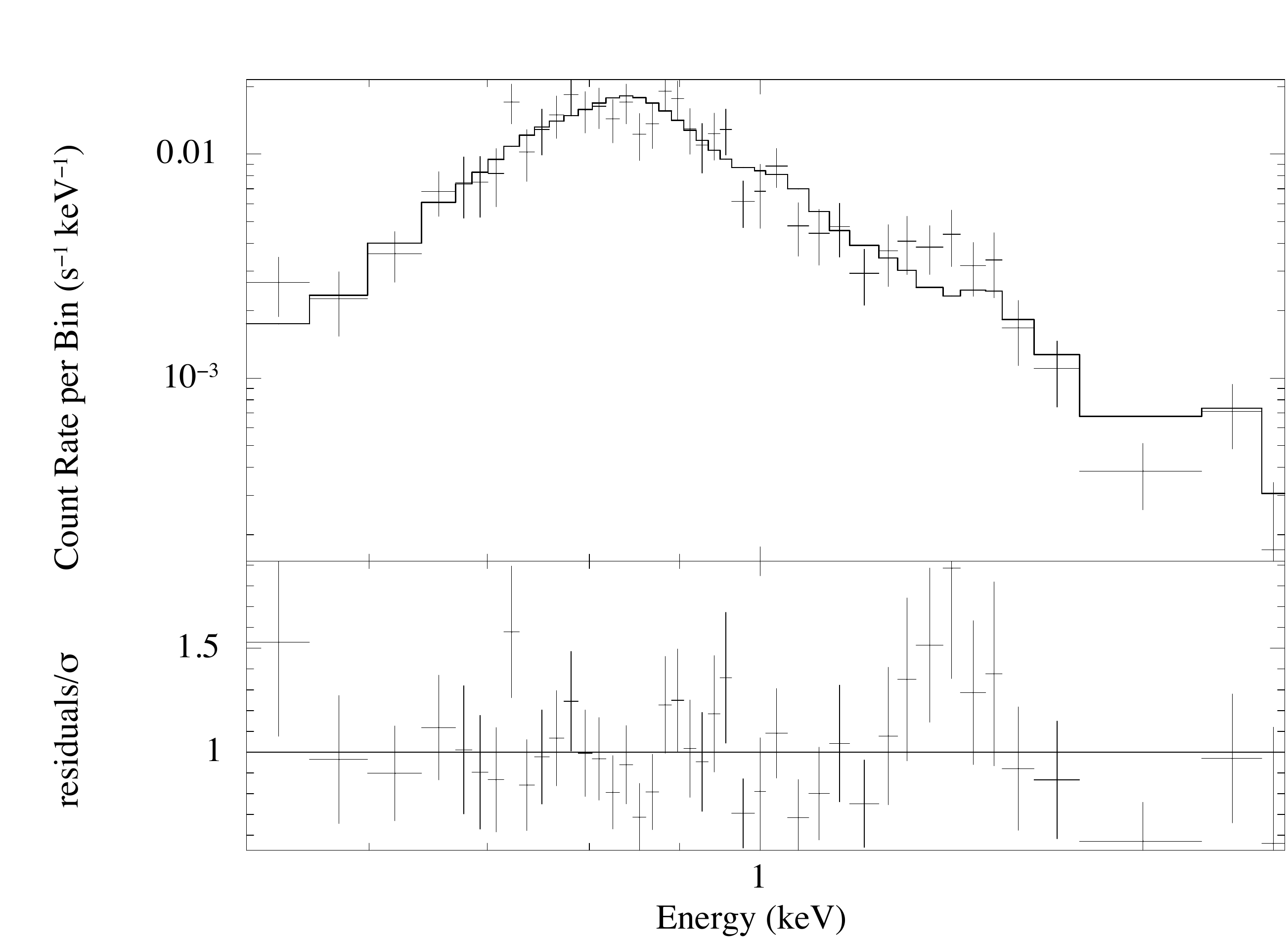} \\
\end{tabular}
\caption{Left: \Chandra image for source 65 (the \AEx source extraction regions are superimposed in red). Right: the corresponding 0.35-2 keV spectrum, binned to 20 counts per bin for display purposes only. The spectrum is best fit with a thermal plasma model, and is consistent with a stellar origin.}
\label{bright_spec}
\end{figure*}

Source 65 (010959.34+354206.3) is the brightest X-ray source in our total NGC~404 X-ray point source catalog, excluding the NGC~404 central engine. It was detected with 708 net counts in the 0.35-8 keV band during Obs ID 12239, however the X-ray source is very soft -- 97\% of the counts were detected at energies below 2 keV. We therefore refine our spectral modeling by considering only the 0.35-8 keV portion of the spectrum. The best fit model is a thermal plasma \texttt{APEC} model, with $kT$=0.57$^{+0.03}_{-0.04}$ keV ($C/dof$=106/98), predicting a  0.35-8 keV flux of 3.6$\times10^{-14}$ \flux. A smoothed image of source 65 from Obs ID 12239 and the 0.35-2 keV spectrum is shown in Figure~\ref{bright_spec}.

Source 65 is spatially coincident within $\sim$0\as6 of a high proper motion F5 star \citep[HD 6892,][]{Hog+00}, with $M_V\sim8.5$ (located at a distance of $d\sim110$ pc). This optical magnitude implies an X-ray-to-optical flux ratio $log(f_X/f_V)$ of -2.9, typical of the X-ray-to-optical flux ratios reported for F stars by \cite{Agueros+09}. A UV counterpart has also been detected by GALEX, with a FUV magnitude of $17.60\pm0.07$ (see \S\ref{uv}). Our observations support the classification of this object as a foreground star not previously detected in X-rays. %(12239 - 7, 6.8 am; 384 - 8, 8.0am)

\section{Multi-wavelength Comparisons}
\subsection{Optical Counterparts from \HST}
We searched for optical counterpart candidates for each of the 11 X-ray sources that was contained within one or more of the overlapping \HST fields. The typical \Chandra error circle for these sources was $\sim$0\as6. We kept only those optical sources that met the quality cuts described in \cite{Williams+10} (i.e., sources with sufficient signal-to-noise, not flagged as unusable, meeting predetermined crowding and sharpness thresholds, etc.). The \HST exposures reach a typical limiting magnitude of 26 in $F814W$ and 27 in $F606W$, equivalent to $M_{814} = -1.4$ and $M_{606} = - 0.4$ (i.e., late B or early A-type main sequence stars).

\begin{figure*}
\centering
\begin{tabular}{cccc}
Source 26 & Source 27 & Source 29 & Source 30 \\
\includegraphics[width=0.23\linewidth,clip=true,trim=2cm 4cm 2cm 4cm]{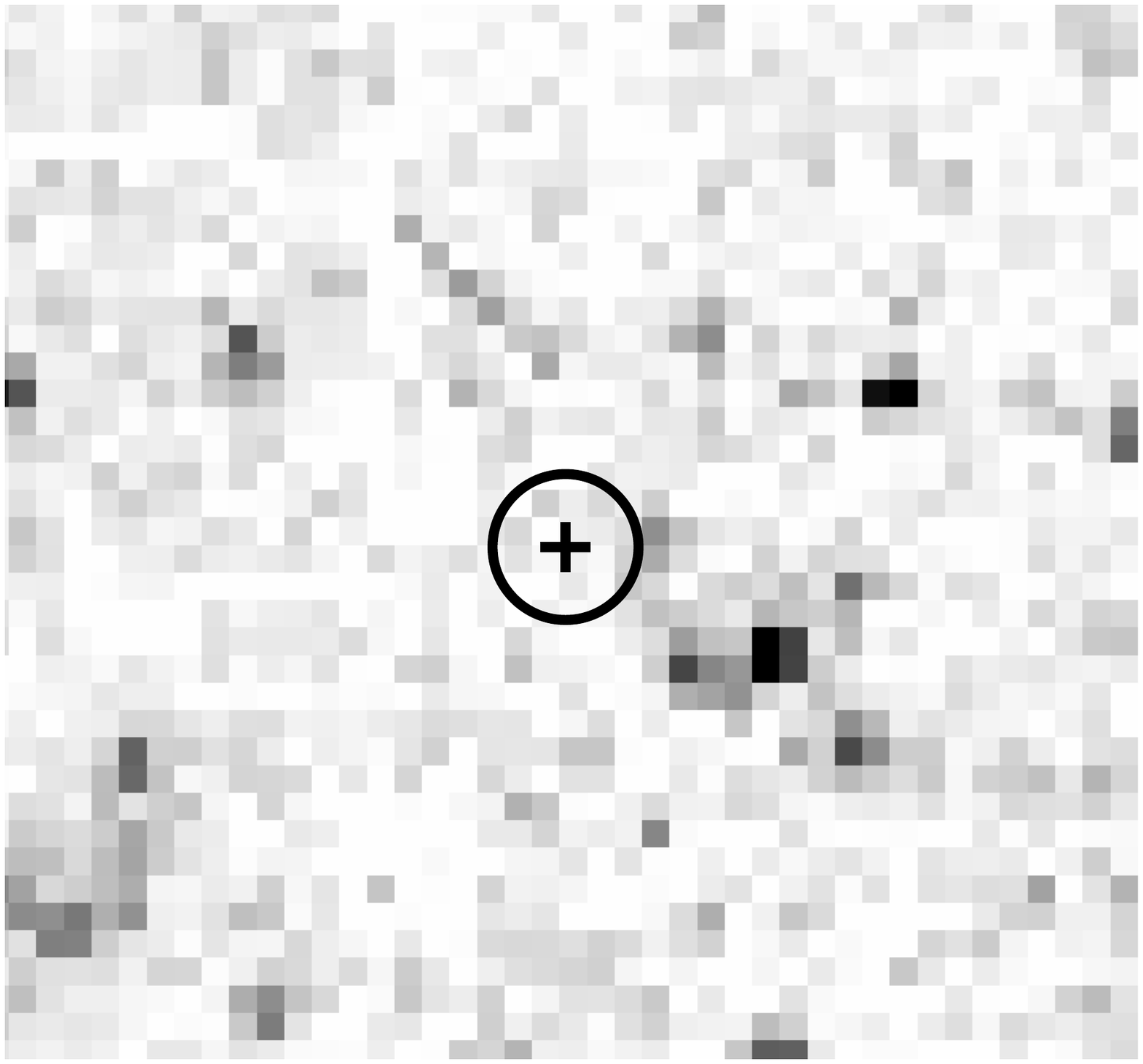} &
\includegraphics[width=0.23\linewidth,clip=true,trim=2cm 4cm 2cm 4cm]{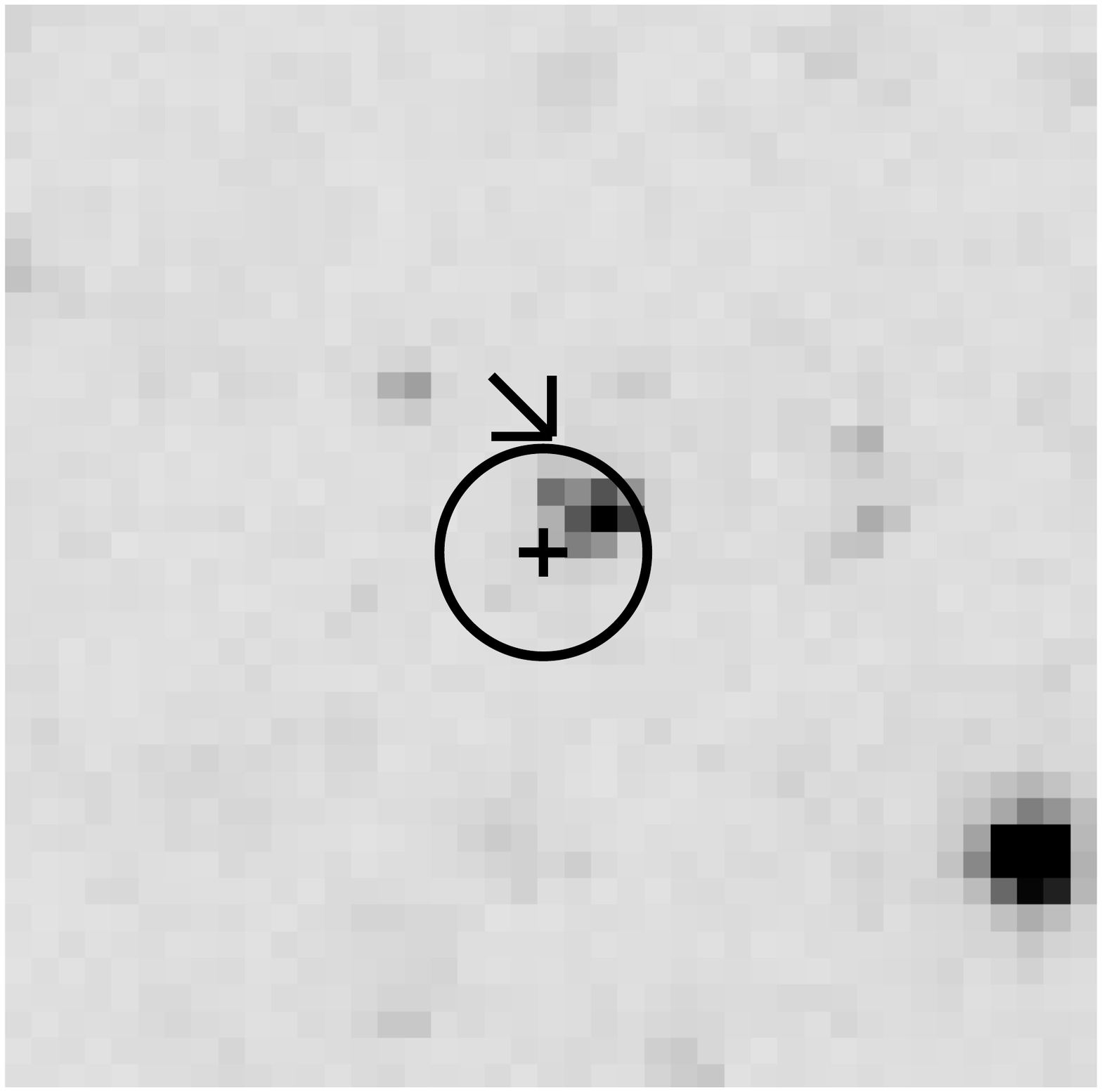} &
\includegraphics[width=0.23\linewidth,clip=true,trim=2cm 4cm 2cm 4cm]{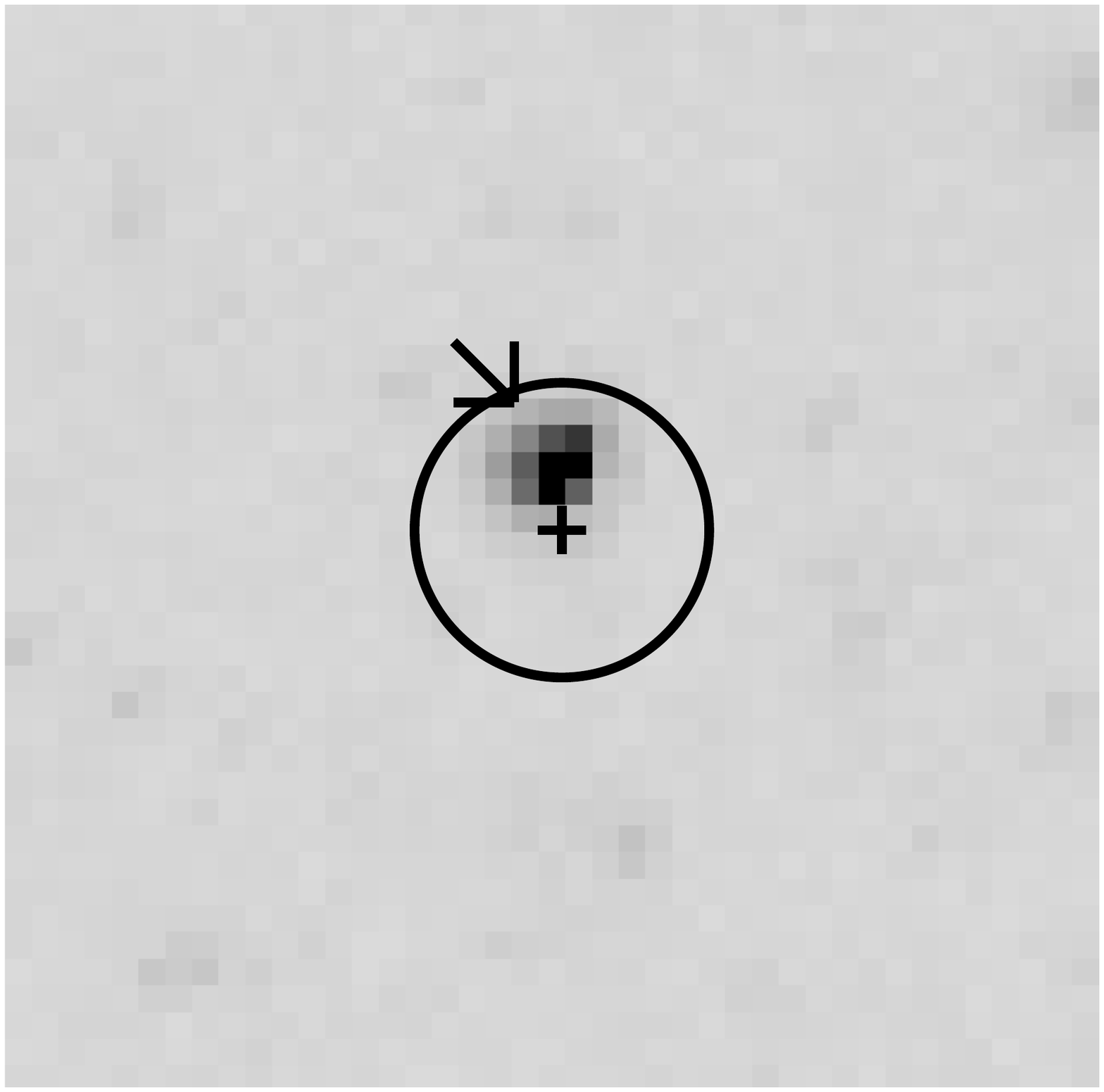} &
\includegraphics[width=0.23\linewidth,clip=true,trim=2cm 4cm 2cm 4cm]{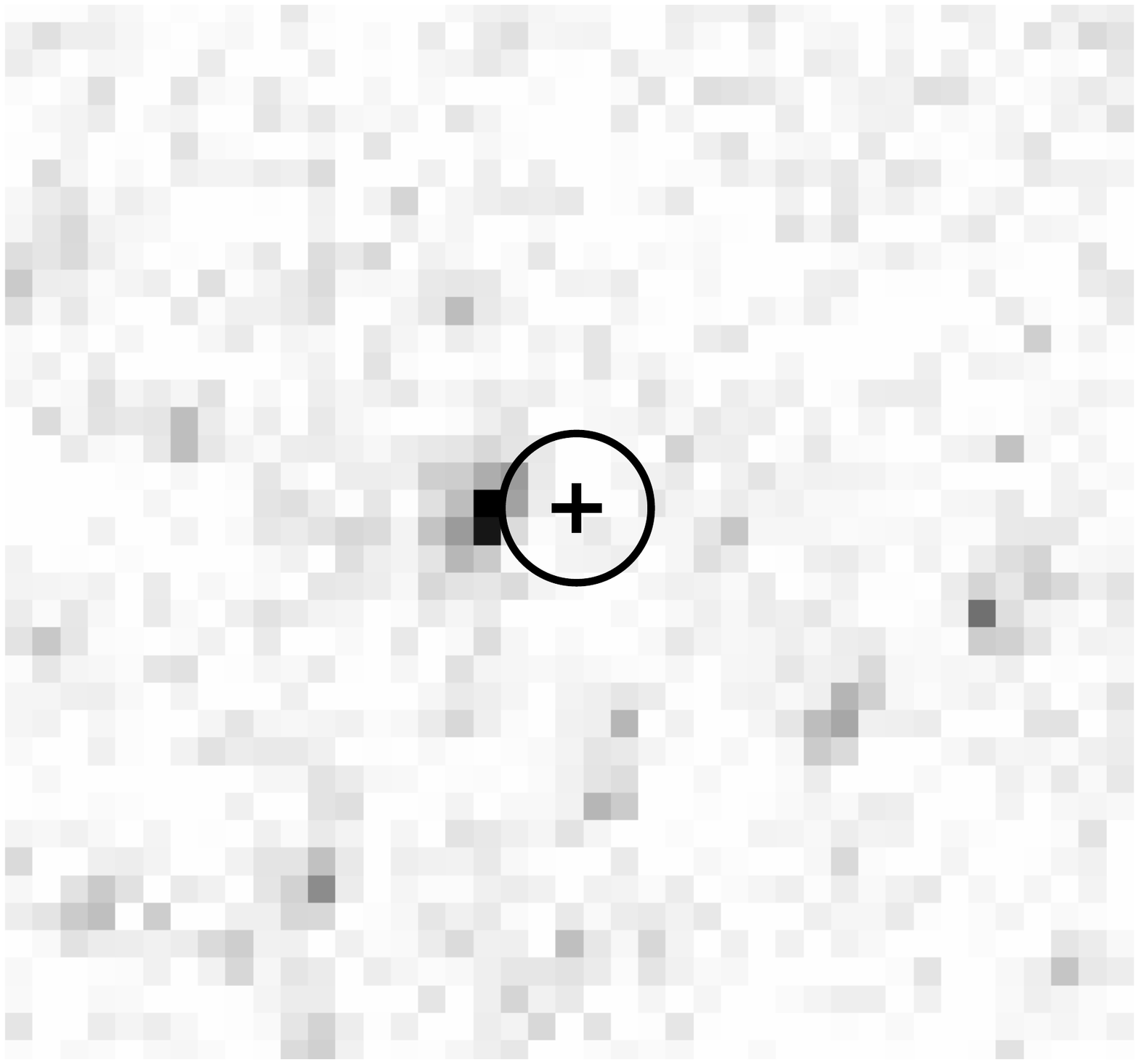} \\
\smallskip \\
Source 34 & Source 35 & Source 39 & Source 40 \\
\includegraphics[width=0.23\linewidth,clip=true,trim=2cm 4cm 2cm 4cm]{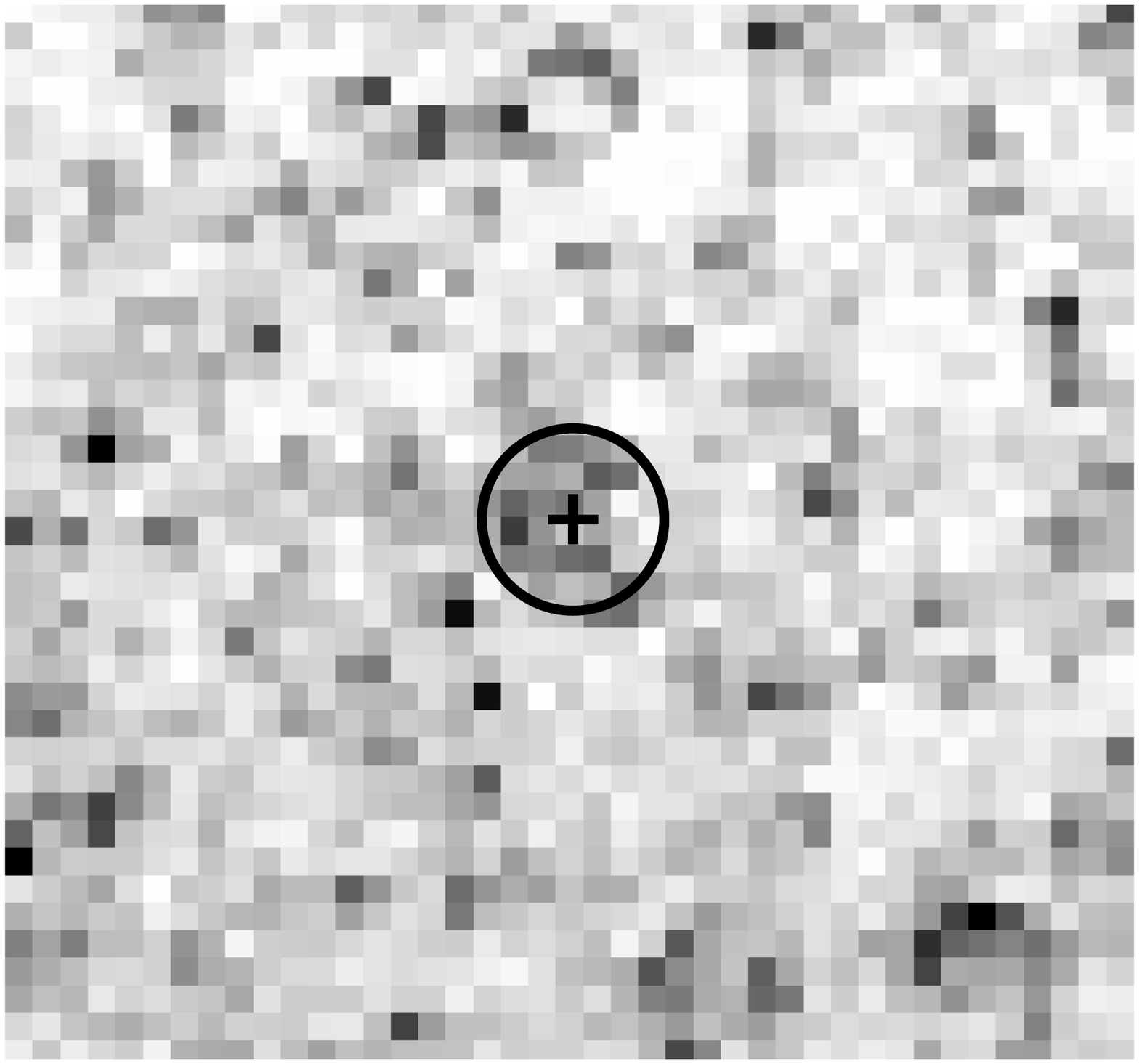} &
\includegraphics[width=0.23\linewidth,clip=true,trim=2cm 4cm 2cm 4cm]{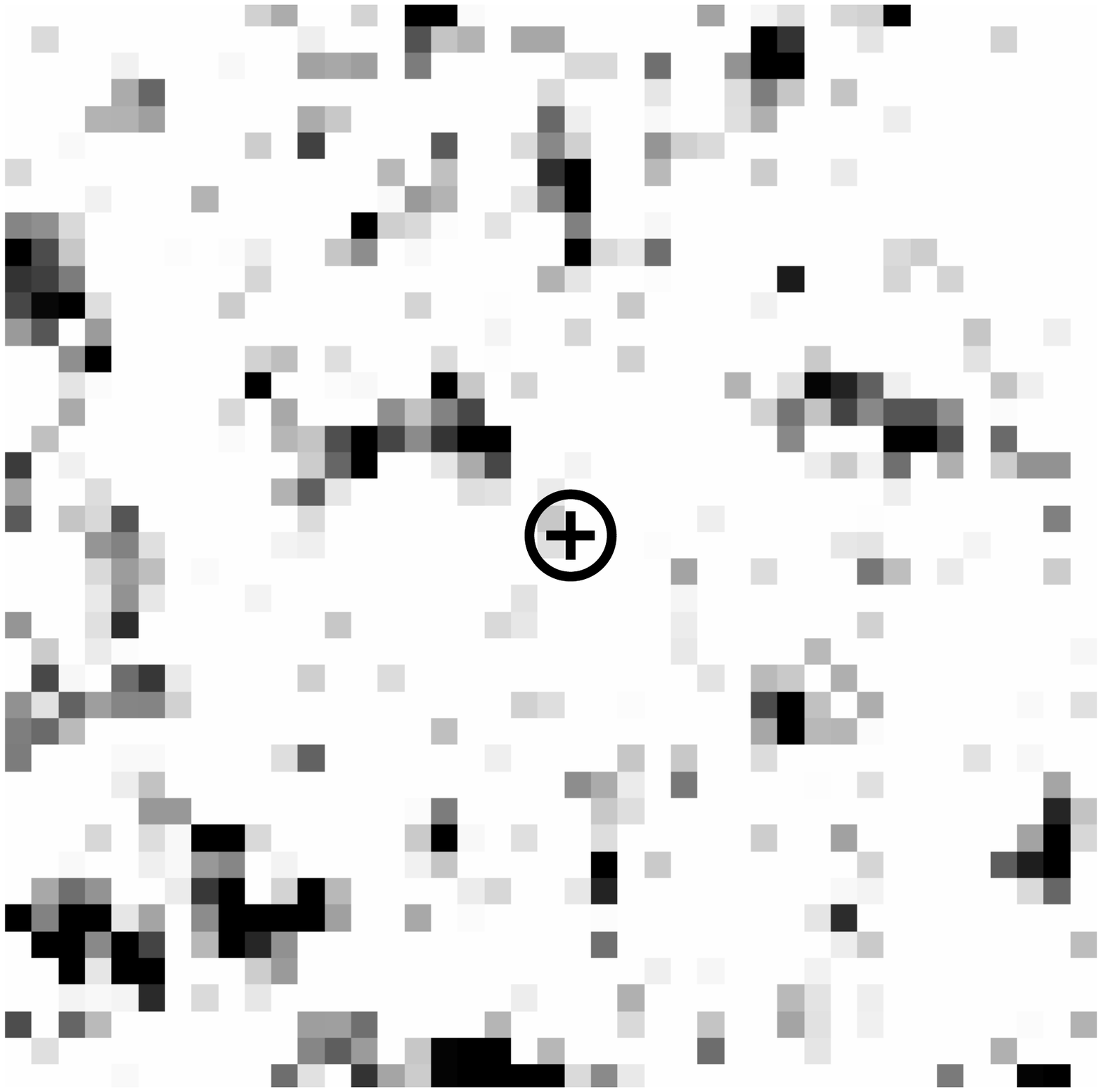} &
\includegraphics[width=0.23\linewidth,clip=true,trim=2cm 4cm 2cm 4cm]{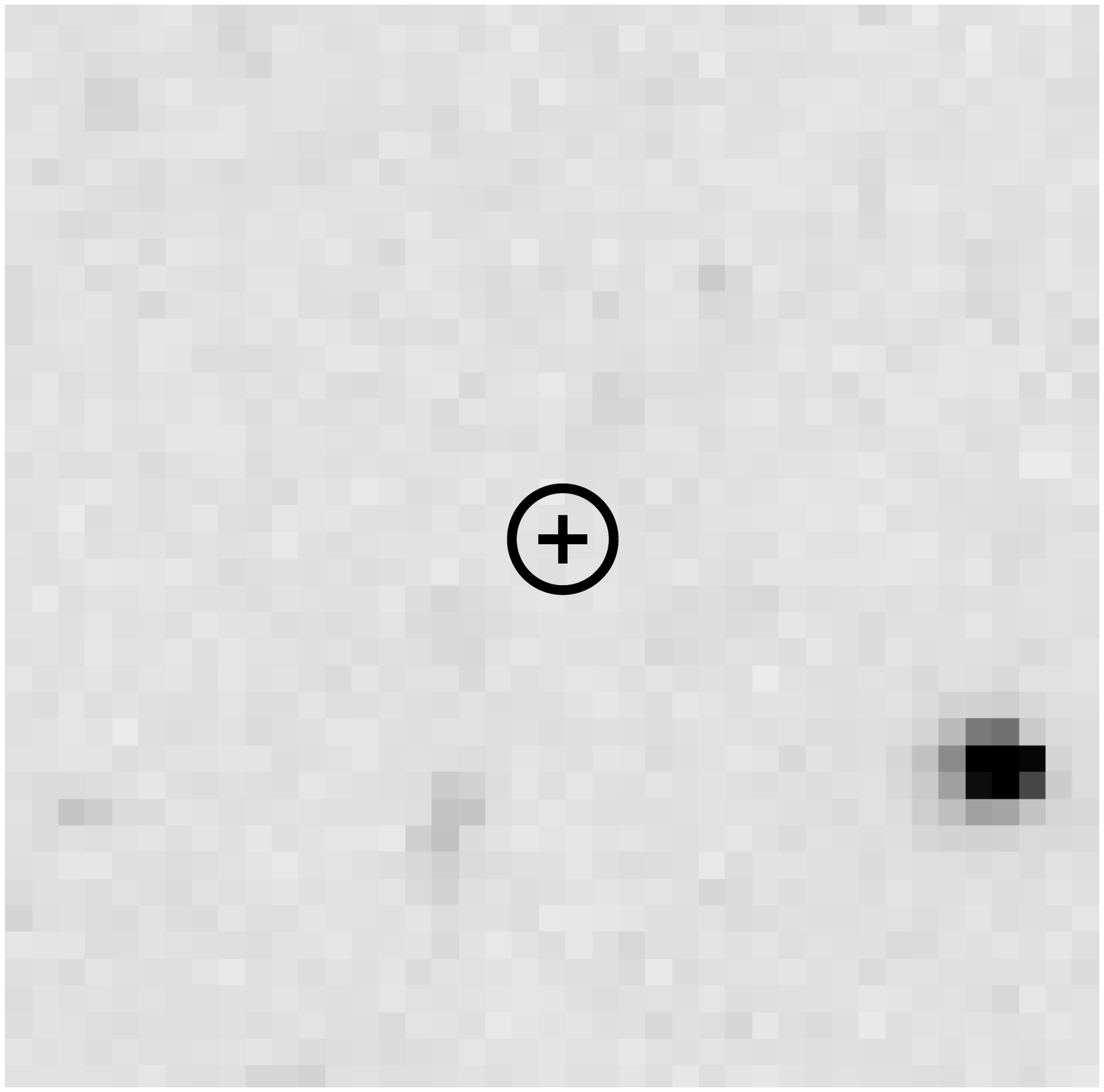} &
\includegraphics[width=0.23\linewidth,clip=true,trim=2cm 4cm 2cm 4cm]{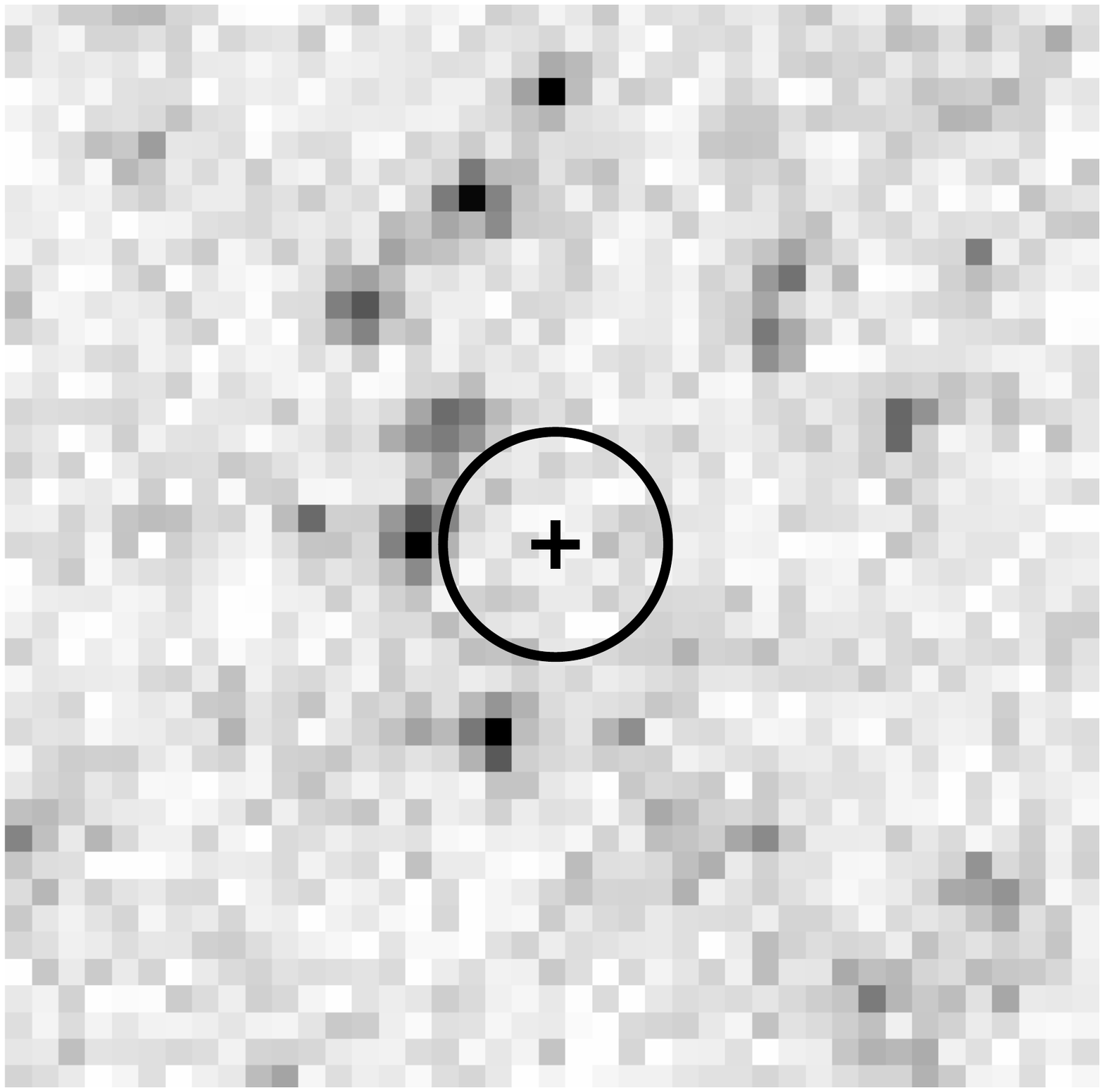} \\
\smallskip \\
Source 42 & Source 44 & Source 50 & \\
\includegraphics[width=0.23\linewidth,clip=true,trim=2cm 4cm 2cm 4cm]{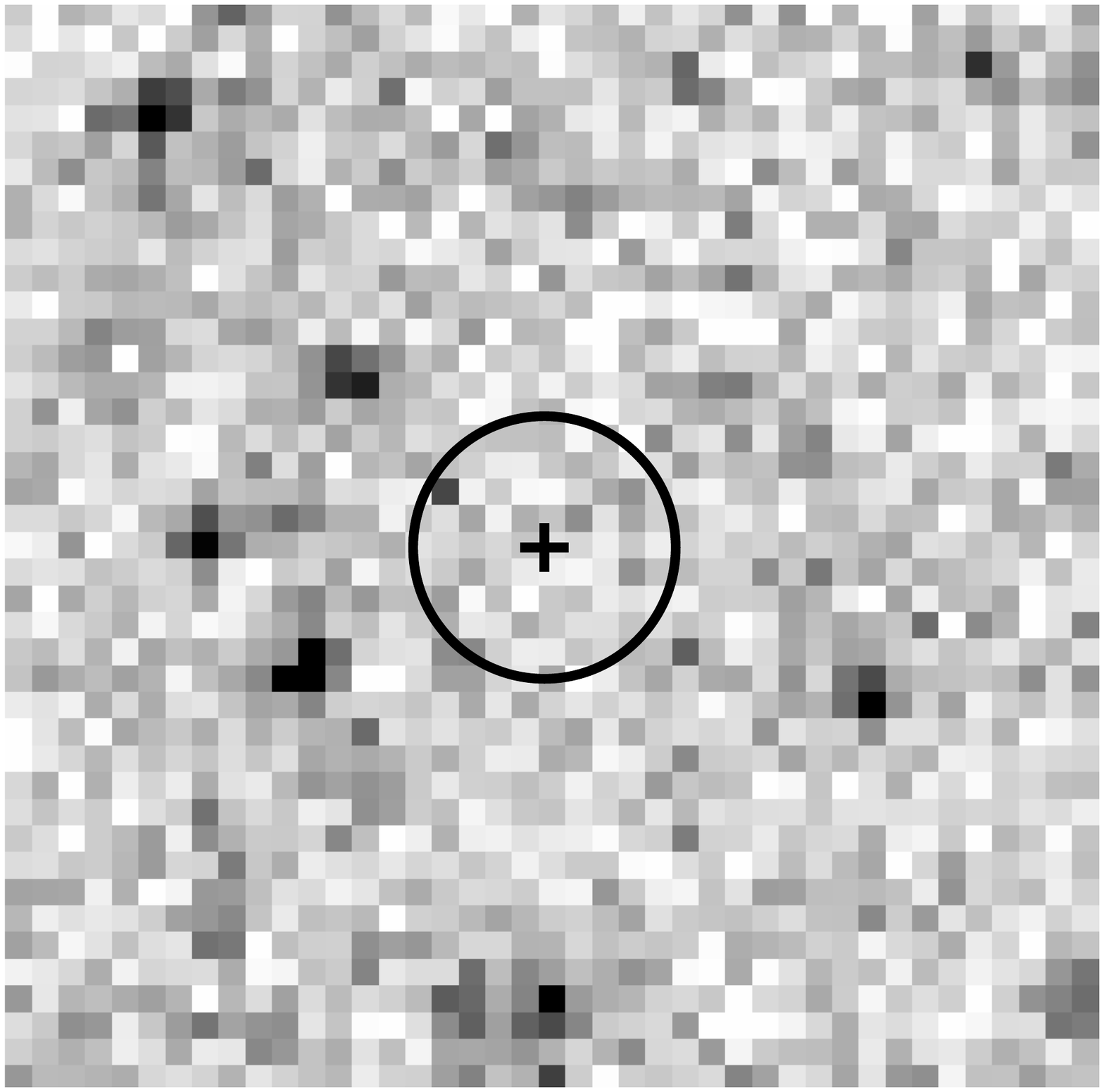} &
\includegraphics[width=0.23\linewidth,clip=true,trim=2cm 4cm 2cm 4cm]{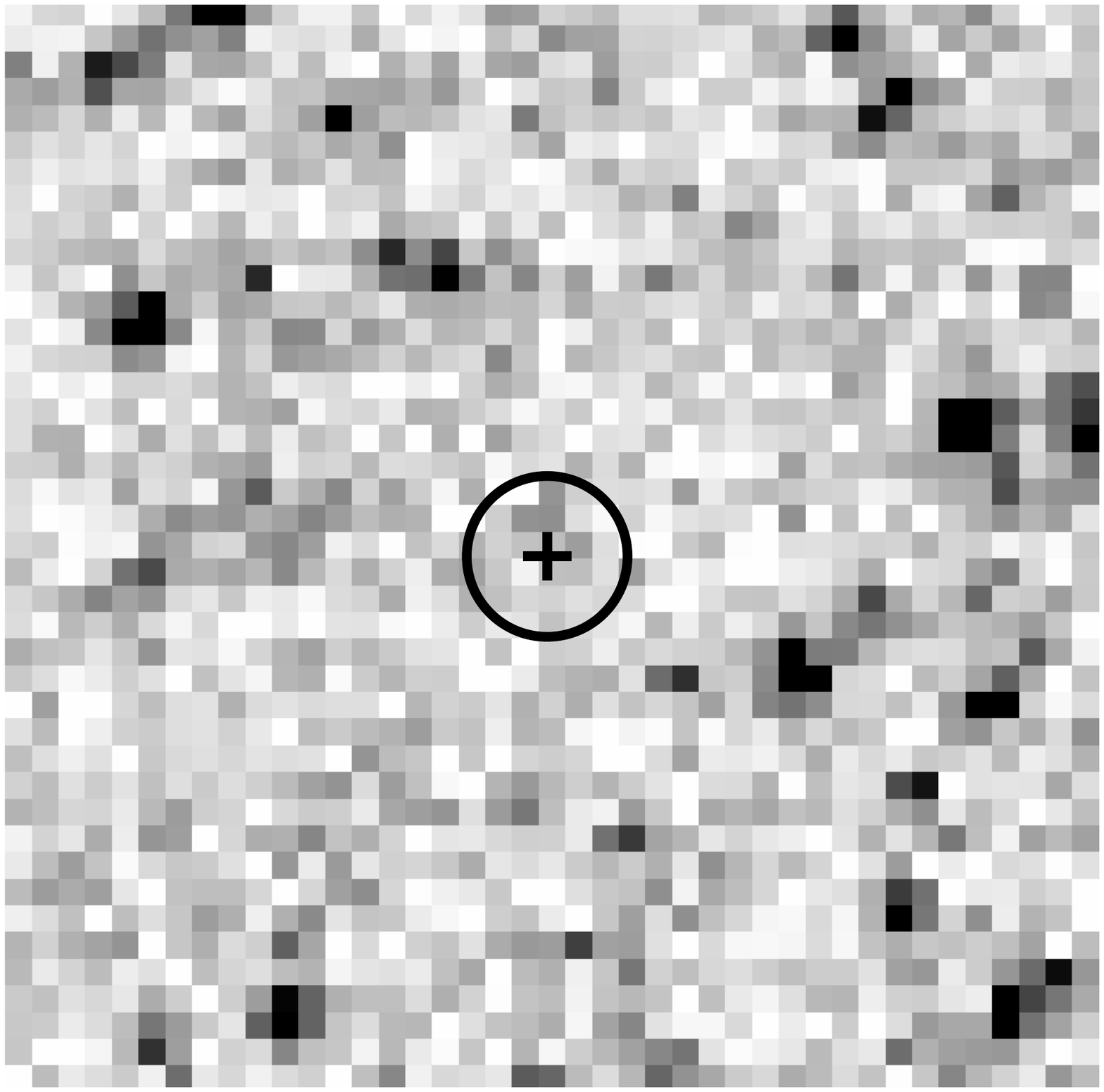} &
\includegraphics[width=0.23\linewidth,clip=true,trim=2cm 4cm 2cm 4cm]{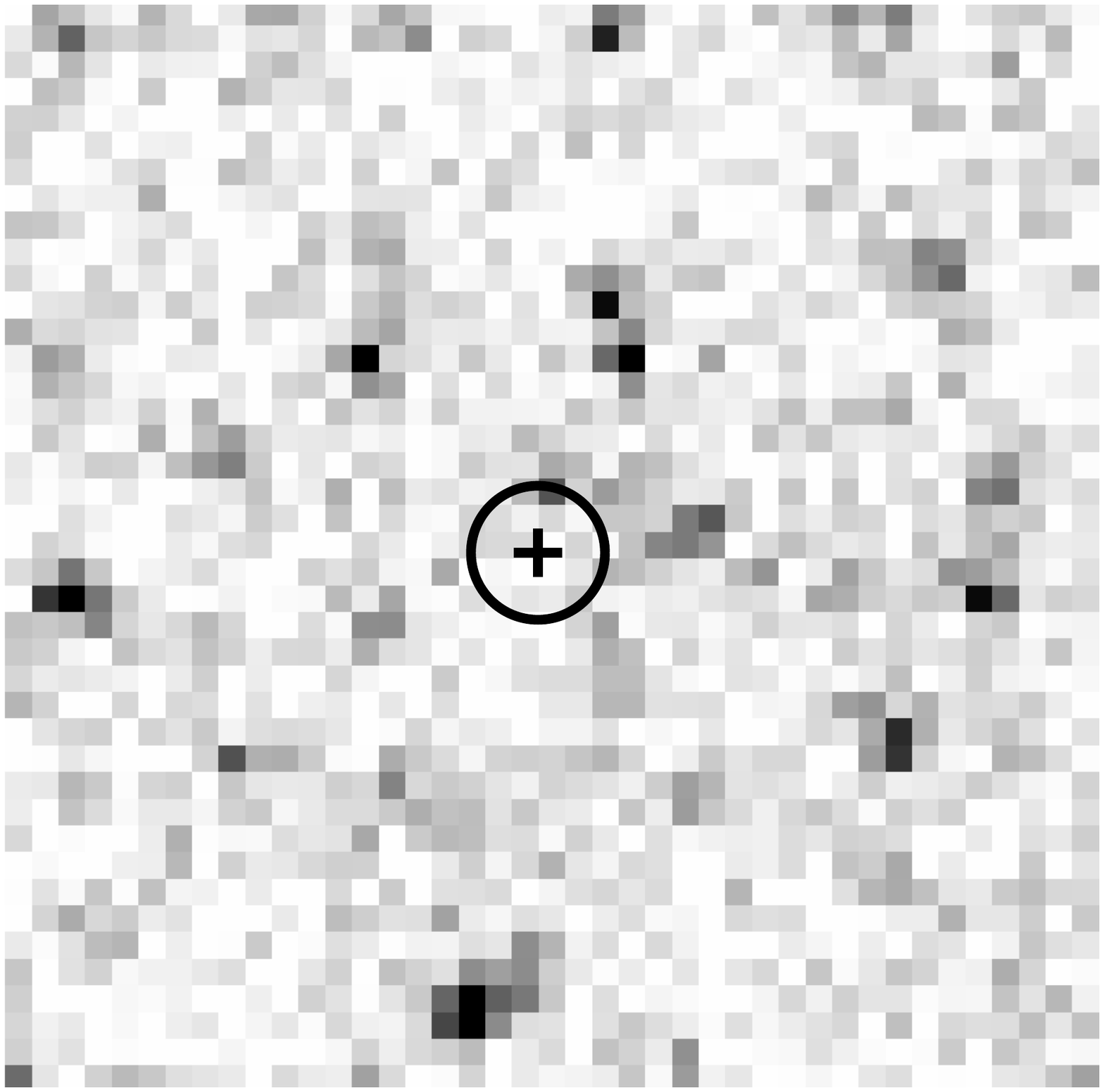} & \\
\end{tabular}
\caption{\HST $F606W$ postage stamps. Each square is 5\asn$\times$ 5\asn. The black cross shows the \Chandra X-ray source position, and the circle denotes the error circle. Optical counterpart candidates are noted with an arrow. Sources 26, 34, 40, and 42 are XRB candidates.}
\label{optical}
\end{figure*}

Figure~\ref{optical} shows $F606W$ 5\asn$\times$5\asn finding charts for the 11 X-ray sources with overlapping \HST fields. Only two X-ray sources (27 and 29) were found to have optical counterpart candidates, shown in Figure~\ref{optical} with arrows. The source 27 optical counterpart has $F606W=25.33\pm0.02$ and $F606W-F814W=1.08\pm0.02$, while the optical counterpart to source 29 has $F606W=21.68\pm0.01$ and $F606W-F814W=2.47\pm0.03$. The optical colors of both sources are consistent with a red giant branch (RGB) star or a background AGN. We estimate the X-ray-to-optical flux ratio $log(f_X/f_V)$ (where $f_X$ is the 2-8 keV flux and $f_V$ is the $F606W$ flux). For source 27, we find $log(f_X/f_V)=0.74$, while source 29 has $log(f_X/f_V)=-2.14$. These X-ray-to-optical flux ratios are consistent with an AGN or LMXB origin. The low number of optical counterparts, and the lack of any blue optical counterpart candidates, suggests that there are few HMXBs in NGC~404.

\subsection{Ultraviolet, Optical, Infrared, and Radio Counterparts}\label{uv}
We have attempted to place additional constraints on the nature of the NGC~404 X-ray sources by examining archival multi-wavelength data. We searched for counterparts in the following point source catalogs: the {\it GALEX} GR6 data release (NUV, FUV), USNO-B1.0 ($B$, $R$), 2MASS All Sky ($J$, $H$, $K$), and NVSS (1.4 GHz). 

UV emission can be a good tracer of young stars and background AGN; however, since the stellar populations of NGC~404 are primarily old, we expect the majority of UV-detect sources to be background AGNs. It is possible that some LMXBs in NGC~404 will exhibit detectable levels of UV emission, although their optical colors are expected to be significantly redder than that of an AGN. The optical USNO-B1.0 catalog is sensitive to Milky Way stars, compact clusters, and AGN. The infrared data are also sensitive to Milky Way stars, red giants, associations of older stars located in NGC~404, and background galaxies, while radio emission is indicative of radio-loud background AGN.

Table~\ref{multiwavelength} summarizes the multi-wavelength observations available for each of the 31 X-ray sources with at lease one multi-wavelength detection. Ten of these sources have counterpart detections in multiple wavelengths.

\begin{sidewaystable*}
\vspace*{8cm}
\centering
\caption{Multi-wavelength Observations of NGC~404 X-ray Sources}
\begin{tabular}{ccccccccccccccc}
\hline \hline
No.	& \multicolumn{2}{c}{{\it GALEX}} && \multicolumn{2}{c}{\HST} & & \multicolumn{2}{c}{USNO-B1.0} & & \multicolumn{3}{c}{2MASS} & & NVSS \\   \cline{2-3} \cline{5-6} \cline{8-9} \cline{11-13}
	& NUV	& FUV		 && $F606W$	& $F814W$	& & $B$	& $R$				& & $J$	& $H$	& $K$	& & $S_{\rm 1.4 GHz}$ (mJy) \\
(1) 	& (2)		 & (3)		 && (4)		&  (5)		& & (6)	& (7)					& & (8)	& (9)		& (10)	& & (11) \\
\hline
1	& $22.07\pm0.22$	& 				& & 				& 				& & 20.22	& 18.57	& & 				& 				&				& & 				\\
3	& $22.96\pm0.15$	& 				& &				&				& &		&		& &				&				&				& &				\\
4	&				& 				& &				&				& & 19.22 & 17.21	& & $16.32\pm0.12$	& $15.89\pm0.18$	& $15.47\pm0.23$	& &				\\
5	&				& 				& &				&				& & 19.45 & 18.73	& &				&				&				& &				\\
7	& $21.74\pm0.04$ 	& $22.86\pm0.09$ 	& &				&				& & 19.05 & 20.77	& &				&				&				& &				\\
8	&				& 				& &				&				& &		&		& &				&				&				& & $53.7\pm2.0$	\\
15	& $22.74\pm0.08$ 	& $23.38\pm0.12$ 	& &				&				& &		&		& &				&				&				& &				\\
17	& $22.11\pm0.10$	& 				& &				&				& & 18.83 &		& &				&				&				& &				\\
19	& $22.06\pm0.07$ 	& $22.20\pm0.09$ 	& & 				&				& &		&		& &				&				&				& &				\\
20	& 				& $22.57\pm0.30$ 	& &				&				& &		&		& &				&				&				& &				\\
21	& 				& $24.05\pm0.26$ 	& &				&				& &		&		& &				&				&				& &				\\
27	& $24.30\pm0.21$ 	& 				& & $25.33\pm0.02$	& $24.25\pm0.01$	& & 20.36 & 19.11	& &				&				&				& &				\\
28	& $23.87\pm0.27$ 	&  				& &				&				& &		&		& &				&				&				& &				\\
29	& 				& 				& & $21.68\pm0.01$	& $19.21\pm0.01$	& &		&		& &				&				&				& &				\\
32	& $21.69\pm0.41$ 	& 				& &				&				& &		&		& &				&				&				& &				\\
39	& $22.14\pm0.11$ 	& 				& &				&				& &		&		& &				&				&				& &				\\
41	& $15.118\pm0.001$ & $17.070\pm0.003$ & &			&				& &		&		& & $11.46\pm0.04$ & $10.56\pm0.03$ 	& $10.55\pm0.04$	& & $3.4\pm0.4$	\\
44	& $20.73\pm0.36$ 	& 				& &				&				& &		&		& &				&				&				& &				\\
50	& $21.94\pm0.38$ 	& 				& &				&				& &		&		& &				&				&				& &				\\
51	& $20.18\pm0.17$ 	& $20.45\pm0.31$ 	& &				&				& & 18.39 & 16.59	& & $16.35\pm0.13$ & $15.61\pm0.13$ 	& $14.98\pm0.14$	& &				\\
52	& $23.03\pm0.12$ 	& $23.83\pm0.23$ 	& & 				&				& & 20.75 &		& &				&				&				& &				\\
53	& $19.98\pm0.04$ 	& 				& &				&				& &		&		& &				&				&				& &				\\
59	& $21.72\pm0.09$ 	&  				& &				&				& &		&		& &				&				&				& &				\\
65	&		 		& $17.60\pm0.07$ 	& & 				&				& & 9.02 	& 8.32	& & $7.66\pm0.02$ 	& $7.52\pm0.02$ 	& $7.47\pm0.02$	& &				\\
66	& $22.71\pm0.11$ 	& $23.39\pm0.22$ 	& &				&				& &		&		& &				&				&				& &				\\
67	& $22.50\pm0.10$ 	& $24.07\pm0.19$ 	& & 				&				& & 20.65 & 18.88	& &				&				&				& &				\\
68	&				& 				& &				&				& & 20.64 &		& &				&				&				& &				\\
69	&				& 				& &				&				& &		&		& & $16.25\pm0.09$ & $15.55\pm0.11$ 	& $14.97\pm0.13$	& &				\\
70	& $22.75\pm0.13$ 	& 				& &				&				& &		&		& &				&				&				& &				\\
72	& $24.41\pm0.52$ 	&  				& &				&				& &		&		& &				&				&				& &				\\
74	& $22.37\pm0.06$	& 				& &				&				& &		&		& &				&				&				& &				\\
\hline \hline
\label{multiwavelength}
\end{tabular}
\end{sidewaystable*}

\subsection{Source Classification}
We combine all of the X-ray and multi-wavelength data available for each source to determine a likely source classification. The first indicator we consider is the source's galactocentric radius; as discussed in Section~\ref{radial_src}, we expect sources with $d>2.5$ kpc to be dominated by contaminants (background AGN and foreground stars), while sources with $d<2.5$ kpc will represent a mix of sources associated with NGC~404 and those that are not. We next take into account the shape of a source's X-ray spectrum using our derived hardness ratios (for faint sources) and X-ray spectral fitting (for sufficiently bright sources). In cases where the spectral fits and hardness ratios conflicted, we only consider the spectral fit. This occurred for only three sources (20, 32, and 45), and in all cases the sources could be described, within the uncertainties, by multiple HR classifications (i.e., source 20 is classified as `SOFT' but the HR uncertainties may also place it within the `XRB' category). Finally, we consider the detection of a multi-wavelength counterpart. 

Sources with $d>2.5$ kpc and hardness ratios `XRB,' `HARD,' `BKG,' `ABS,' or with an X-ray spectrum best fit by a power law are classified as AGNs. Those sources with additional UV or radio detections are highly likely to be background AGNs. X-ray sources with $d>2.5$ kpc and `SOFT,' `SNR,' or thermal spectra are classified as foreground stars (FSs), while softer/thermal sources with $d<2.5$ kpc are potentially super-soft sources (SSSs) and may be associated with NGC~404. We feel especially confident that softer sources with an IR detection in 2MASS are indeed FSs. Sources with $d<2.5$ kpc with `XRB' hardness ratios or a power law spectrum are classified as XRBs. 

Again, `SOFT' and `SNR' sources are classified as FSs, and sources with harder spectra are classified as AGNs. Table~\ref{classifications} lists our final source classifications.

We find 21 candidate XRBs, 46 likely background AGN, 4 candidates SSSs, and 2 likely FSs (plus the NGC~404 central engine). Six sources have been tentatively labeled as `AGN?,' but have at least one source property (for example, $d<$2.5 kpc) that makes them a potential XRB candidate. Given the \cite{Cappelluti+09} AGN \lognlogs distribution, we expect $\sim37-40$ X-ray sources to be background AGN; it is therefore likely that at least some of these six AGN candidates are in fact associated with NGC~404. We find no obvious HMXB candidates in our X-ray source catalog.

\begin{table*}[ht]
\centering
\caption{X-ray Source Properties and Classification}
\begin{tabular}{cccccccc}
\hline \hline
No.	& Class		& Spectral			& Variability	& $d$ 	& Counterparts? 	& Classification 	\\
	& from HRs	& Fit				&			& (kpc)	& 				& 				\\
(1) 	& (2)			& (3)  			& (4)			&  (5)	& (6)				& (7)				 \\
\hline
1	& XRB		& pow, \PL=1.65	& rapid		& 4.8		& UV, optical		& AGN \\
2	& XRB		& pow, \PL=0.6		&			& 4.5		& none			& AGN \\
3	& XRB		& pow, \PL=1.5		&			& 4.3		& UV				& AGN \\
4	& XRB		&				&			& 4.1		& optical, IR		& AGN \\
5	& SOFT		&				&			& 3.8		& optical			& FS \\
6	& XRB		&				&			& 3.8		& none			& AGN \\
7 	& XRB 		& pow, \PL=1.8 	&			& 4.1 	& UV, optical		& AGN \\
8	& XRB		&				&			& 4.1		& radio			& AGN \\ 
9	& XRB		&				&			& 3.3		& none			& AGN \\
10	& ABS		&				&			& 3.1		& none			& AGN \\
11	& ABS		& pow, \PL=0.8		&			& 3.5		& none			& AGN \\
12	& ABS		&				&			& 2.7		& none			& AGN \\
13	& XRB		& pow, \PL=1.2		&			& 3.4		& none			& AGN \\
14	& XRB		&				&			& 2.6		& none			& AGN \\
15 	& HARD 		&				&			& 3.3 	& UV				& AGN \\
16	& ABS		&				& long		& 2.4		& none			& XRB \\
17	& XRB		& pow, \PL=1.9		&			& 2.0		& UV, optical		& AGN \\
18	& XRB		&				& long		& 2.0		& none			& AGN \\
19 	& XRB 		& 				& 			& 2.7 	& UV				& AGN \\
20	& XRB		& pow, \PL=1.5		&			& 1.7		& UV				& AGN? \\
21	& XRB		&				&			& 1.7		& UV				& AGN? \\
22	& SOFT		&				&			& 2.4		& none			& XRB \\
23	& ABS		&				&			& 1.6		& none			& AGN \\
24	& ABS		&				&			& 2.0		& none			& AGN \\
25	& ABS		&				&			& 1.5		& none			& AGN \\
26	& XRB		&				&			& 1.4		& none			& XRB \\
27	& ABS 		&				& long?		& 1.7 	& UV, \HST, optical	& AGN? \\
28	& XRB		&				&			& 1.9		& UV				& AGN? \\
29	& SNR 		& 				& 			& 1.2 	& UV, \HST		& SSS? \\
30	& XRB		&				&			& 1.2		& none			& AGN \\
31	& SOFT		&				&			& 1.3		& none			& AGN \\
32	& XRB		& apec, $kT$=0.8	&			& 1.7		& UV				& SSS? \\
33	& XRB		&				&			& 1.9		& none			& AGN \\
34	& HARD		&				&			& 0.6		& none			& XRB \\
35	& ABS		& abs. apec, $kT$=2.8 &			& 1.6		& none			& SSS? \\
36	& HARD		&				&			& 0.6		& none			& AGN \\
37	& HARD		&				& long		& 0.4		& none			& XRB \\
38	& XRB		&				&			& 0.3		& none			& XRB \\
39	& SOFT		& pow, \PL=1.4		&			& 1.4		& UV				& AGN? \\
40	& ABS		&				& long		& 1.1		& none			& XRB \\
41	& XRB 		& pow, \PL=2.1 	& 			& 0.0 	& UV, IR, radio		& NUCLEUS \\
42	& XRB		&				&			& 1.5		& none			& XRB \\
43	& SNR		&				&			& 0.1		& none			& SSS? \\
44	& ABS		&				&			& 1.1		& UV				& AGN \\
45	& SOFT		& pow, \PL=1.6		&			& 1.3		& none			& AGN \\
46	& XRB		& pow, \PL=1.4		&			& 1.7		& none			& AGN \\
47	& XRB		&				&			& 0.6		& none			& XRB \\
48	& XRB		&				&			& 0.6		& none			& XRB \\
49	& XRB		&				& long?		& 0.5		& none			& XRB \\
50	& ABS		&				&			& 1.2		& UV				& AGN? \\
51	& XRB 		& 				& 			& 1.6 	& UV, optical, IR	& XRB \\
52	& XRB 		& 				& 			& 1.9 	& UV, optical		& XRB \\
53	& SOFT		&				& long		& 0.9		& UV				& XRB? \\
54	& XRB		&				& long		& 1.0		& none			& AGN \\
55	& XRB		& pow, \PL=0.9		& long		& 1.7		& none			& XRB? \\
56	& XRB		&				&			& 1.5		& none			& XRB \\
57	& XRB		&				& long?		& 1.7		& none			& XRB \\
58	& SOFT		& pow, \PL=1.5		&			& 2.1		& none			& XRB \\
59	& XRB		&				& long		& 1.9		& UV				& XRB \\
60	& ABS		&				&			& 2.5		& none			& XRB \\
\hline \hline
\label{classifications}
\end{tabular}
\end{table*}

\setcounter{table}{12}
\begin{table*}[ht]
\centering
\caption{X-ray Source Properties and Classification ({\it continued})}
\begin{tabular}{cccccccc}
\hline \hline
No.	& Class		& Spectral			& Variability	& $d$ 	& Counterparts? 	& Classification 	\\
	& from HRs	& Fit				&			& (kpc)	& 				& 				\\
(1) 	& (2)			& (3)  			& (4)			&  (5)	& (6)				& (7)				 \\
\hline
61	& XRB		&				& long		& 2.5		& none			& XRB \\
62	& ABS		&				&			& 2.6		& none			& AGN \\
63	& XRB		&				&			& 3.1		& none			& AGN \\
64	& XRB		& pow, \PL=1.8		& long		& 3.2		& none			& AGN \\
65	& SNR		& apec, $kT$=0.6	&			& 3.1		& UV, optical, IR	& FS \\
66	& XRB 		& 				& 			& 3.8 	& UV				& AGN \\
67	& XRB 		& pow, \PL=1.2 	& 			& 4.2 	& UV, optical		& AGN \\
68	& XRB		& pow, \PL=1.2		&			& 4.6		& optical			& AGN \\
69	& SOFT		&				&			& 3.9		& IR				& AGN \\
70	& XRB		& pow, \PL=1.8		&			& 4.0		& UV				& AGN \\
71	& XRB		&				& long		& 4.8		& none			& AGN \\
72	& XRB		& pow, \PL=1.7		&			& 4.8		& UV				& AGN \\
73	& ABS		& pow, \PL=1.1		& rapid		& 4.9		& none			& AGN \\
74	& XRB		&				&			& 6.3		& UV				& AGN \\
\hline \hline
\label{classifications}
\end{tabular}
\end{table*}

\subsection{Comparison to Other X-ray Source Populations}
While HMXBs are expected to trace the SFR, LMXBs have been shown to correlate with the stellar mass of the host galaxy. We estimate the number ratio of LMXBs to HMXBs in NGC~404 \citep{Gilfanov04} assuming a stellar mass of NGC~404 of $6.9\times10^8$ \Msun \citep{Thilker+10}; LMXBs are expected to outnumber HMXBs by a factor of 4/10/18 at luminosities greater than 10$^{35}$/10$^{36}$/10$^{37}$ \lum. It is therefore possible that 1-2 of the X-ray sources in our catalog are HMXBs associated with NGC~404. Although we do not observe a high mass optical counterpart to an X-ray source, we note that less than 15\% of the X-ray sources detected by \Chandra have overlapping \HST coverage.

The number of X-ray sources normalized by stellar mass was investigated by \cite{Gilfanov04} for both early and late type galaxies. The galaxies included in the \cite{Gilfanov04} sample all have stellar masses greater than a few 10$^{10}$ \Msun -- more than two orders of magnitude greater than NGC~404. We therefore test whether NGC~404 follows the same scaling relations between number of LMXBs and stellar mass as its more massive counterparts. We note that the \cite{Gilfanov04} work relied on X-ray source catalogs compiled from the literature; the original publications used different (broad) energy bands, assumed different spectral models for converting between count rates to energy fluxes, and employed different source detection strategies. However, it was anticipated that such differences contributed uncertainties of only $\sim$20-30\%. We assume the 0.35-8 keV energy flux and our standard model for all following analysis.

The \cite{Gilfanov04} work evaluated the number of luminous X-ray sources per unit stellar mass of the host galaxy ($N_X/M_*$) found in late-type galaxies to the populations of early type galaxies (see their Table 5), although no significant difference was found between earlier- and later-type galaxies. The average number of X-ray sources more luminous than 10$^{37}$ \lum per 10$^{11}$ \Msun stellar mass of the host galaxy was found to be 166.3$\pm$20.5. We do not find any of the NGC~404 XRB candidates to have a 0.35-8 keV luminosity above 10$^{37}$ \lum. To find an upper limit on the number of XRBs in NGC~404, we employ a similar strategy as described in Section~\ref{sXLF} for our XLF fitting. We randomly add 10 sources from our list of background AGNs and find the number of sources above 10$^{37}$ \lum, as well as their collective luminosity. This was done 500 times. On average, 2.2$\pm$1.3 sources have a luminosity above 10$^{37}$ \lum with a collective luminosity of $\sim$2.6$\times10^{37}$ \lum. We therefore estimate the number of XRBs more luminous than 10$^{37}$ \lum in NGC~404 to be $<3.5$.

\begin{figure*}
\centering
\begin{tabular}{cc}
\includegraphics[width=0.5\linewidth,angle=0,clip=true,trim=2cm 12.5cm 2cm 2cm]{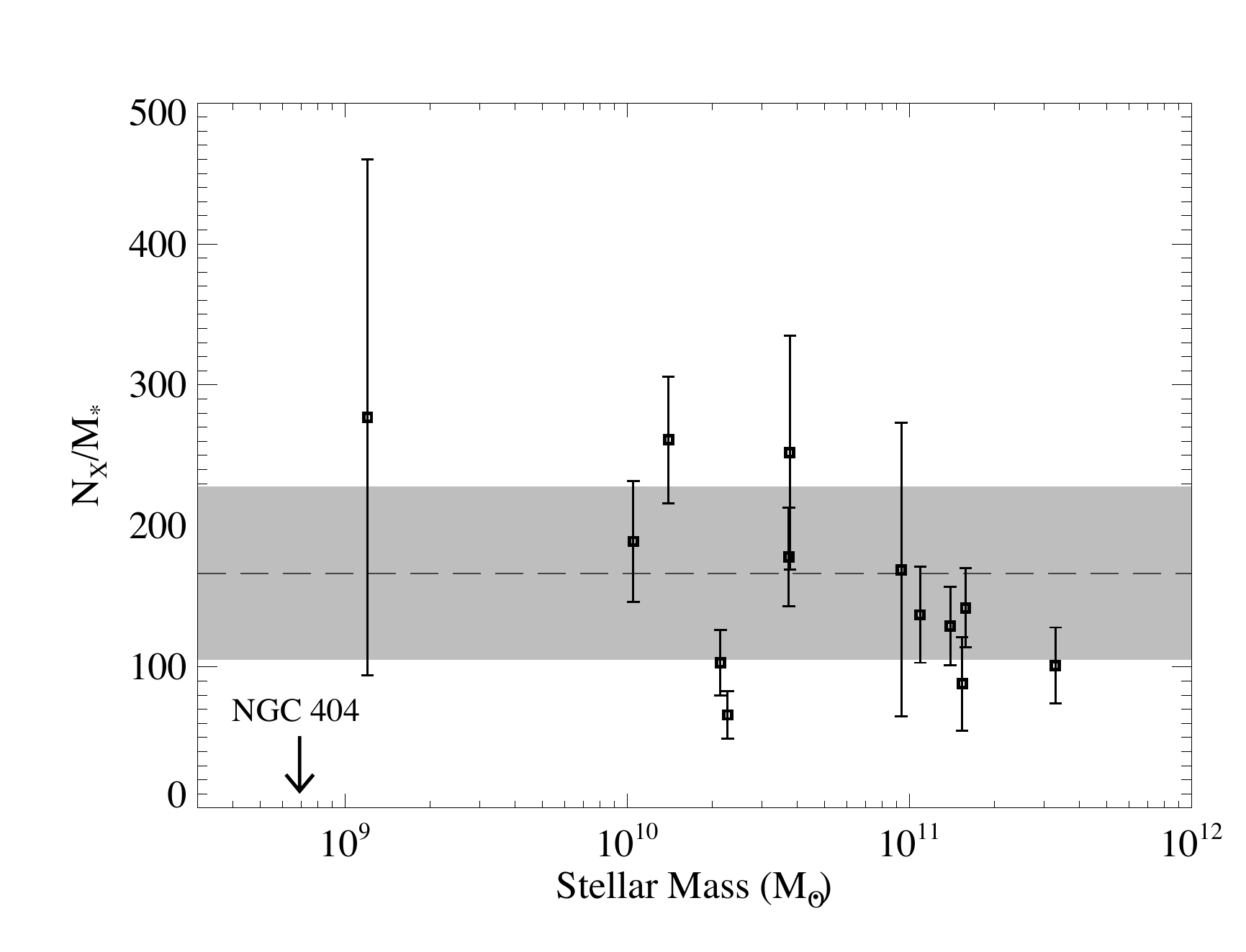}  &
\includegraphics[width=0.5\linewidth,angle=0,clip=true,trim=2cm 12.5cm 2cm 2cm]{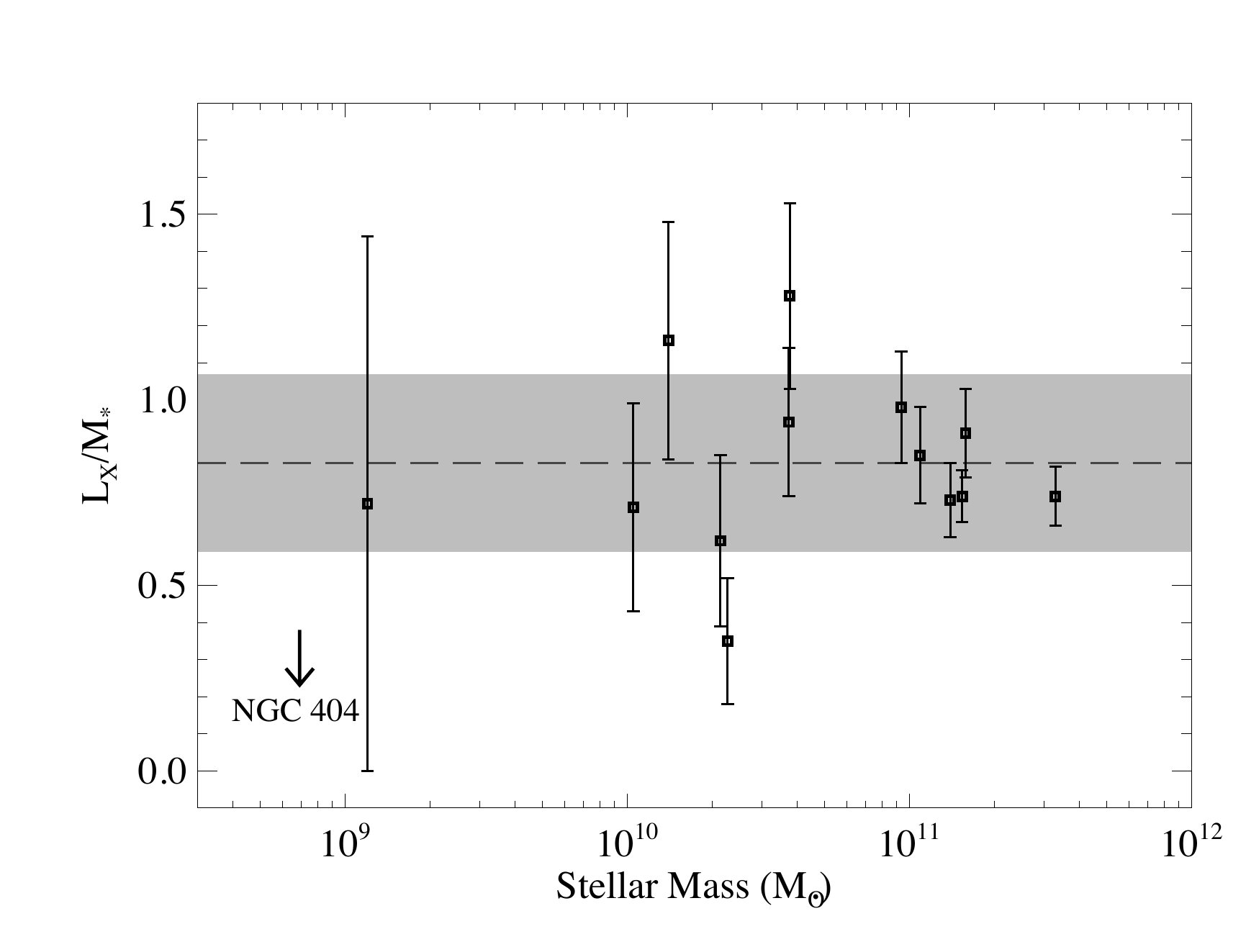}  \\
\end{tabular}
\caption{The left panel shows the number of XRBs more luminous than 10$^{37}$ \lum per unit stellar mass of the host galaxy (in units of 10$^{11}$ \Msun), while the right panel shows the collective luminosity per stellar mass of these XRBs. In both panels, squares show the \cite{Gilfanov04} sample of galaxies, while NGC~404 is shown as an upper limit. The horizontal dashed lines show the average value derived from the \cite{Gilfanov04} sample, and the shaded region shows the 3$\sigma$ uncertainty.}
\label{gilfanov}
\end{figure*}

Assuming NGC~404 hosts $<$3.5 X-ray sources above 10$^{37}$ \lum, and using the same units as \cite{Gilfanov04}, we find $N_X/M_*$ for NGC~404 is $<51$ sources per $10^{10}$ \Msun, more than 5$\sigma$ below the average found for more massive galaxies. Similarly, we compare the collective luminosity of sources above 10$^{37}$ \lum per unit stellar mass of the host galaxy ($L_X/M_*$) of NGC~404 to the more massive galaxies considered in \cite{Gilfanov04}. The sample average from \cite{Gilfanov04} was 0.83$\pm$0.08, where $L_X$ is in units of 10$^{40}$ \lum and $M_*$ is in units of 10$^{11}$ \Msun. In these units, we find an upper limit on $L_X/M_*$ for NGC~404 of 0.38. Although this result is again more than 5$\sigma$ below the average for all galaxies, the $L_X/M_*$ values show considerably more scatter for \cite{Gilfanov04} sample. Figure~\ref{gilfanov} shows both $N_X/M_*$ and $L_X/M_*$ as a function of stellar mass for the \cite{Gilfanov04} sample and NGC~404.

Although there is large amount of scatter among the galaxies considered in \cite{Gilfanov04}, the upper limits on the number of XRBs and the collective luminosity of XRBs per stellar mass found for NGC~404 is low. No obvious trends in these quantities with stellar mass are observed in Figure~\ref{gilfanov}, so it is unclear whether the lower stellar mass of NGC~404 is responsible for this difference. Instead, the relatively recent increase in SFR experienced by NGC~404 may have produced a population of low-luminosity LMXBs powered by MS donor stars which is currently dominating the X-ray emission from XRBs. Future work will be required to understand whether this difference in collective XRB luminosity is common for dwarf galaxies, or if NGC 404 is an outlier.

\section{The X-ray Luminosity Function}\label{sXLF}
Numerous studies have shown a correlation between the shape of the XLF for a given X-ray source population and the SFH of the host galaxy \citep{Kilgard+02,Grimm+03,Gilfanov04,Eracleous+06,Fabbiano06}. We provided a direct observational test of these correlations for young stellar populations in Paper I, where HMXBs were the dominant source of X-ray emission. We now aim to test the XLF slope-age correlation for the older stellar populations found in NGC~404, where LMXBs are expected to be the prominent source of X-rays.

For a survey covering a total geometric area $A$, the cumulative number of sources $N(>S)$ can be evaluated by summing over all sources with fluxes exceeding $S$, weighted by the survey area, $A(S)$, over which a source with flux $S$ could have been detected:

\begin{equation}
N(>S) = \sum\limits_{i,S_i>S}\frac{1}{A(S_i)}.
\end{equation}

Here, $S$ is the energy flux in units of \flux, and $A(S)$ survey area over which a source with flux $S$ could have been detected; $N(>S)$ therefore has units of sources deg$^{-2}$. The $A(S)$ factor accounts for the survey completeness for sources distributed uniformly over the survey area. Although we note that sources associated with NGC~404 may not be uniformly distributed, the radial source distributions from \S~\ref{radial_src} did not reveal any major features in the source distributions, so uncertainties in the completeness over the survey area are likely to be small (less than $\sim10$\%). As a practical matter, we terminate the low flux end of the distribution at the 90\% completeness luminosity of each energy band considered (a few 10$^{35}$ \lum). The sensitivity maps described above allow the area function, $A(S)$, to be evaluated by summing the sensitivity map over the regions where a source with flux $S$ would be detectable. If $N(>S)$ is given, the differential \lognlogs distribution can be computed:

\begin{equation}
N(S)\Delta S = \frac{N(>[S+\Delta S]) - N(>S)}{\Delta S}.
\end{equation}

\begin{figure*}
\centering
\begin{tabular}{cc}
\includegraphics[width=0.5\linewidth,angle=0,clip=true,trim=2cm 12.5cm 2cm 2cm]{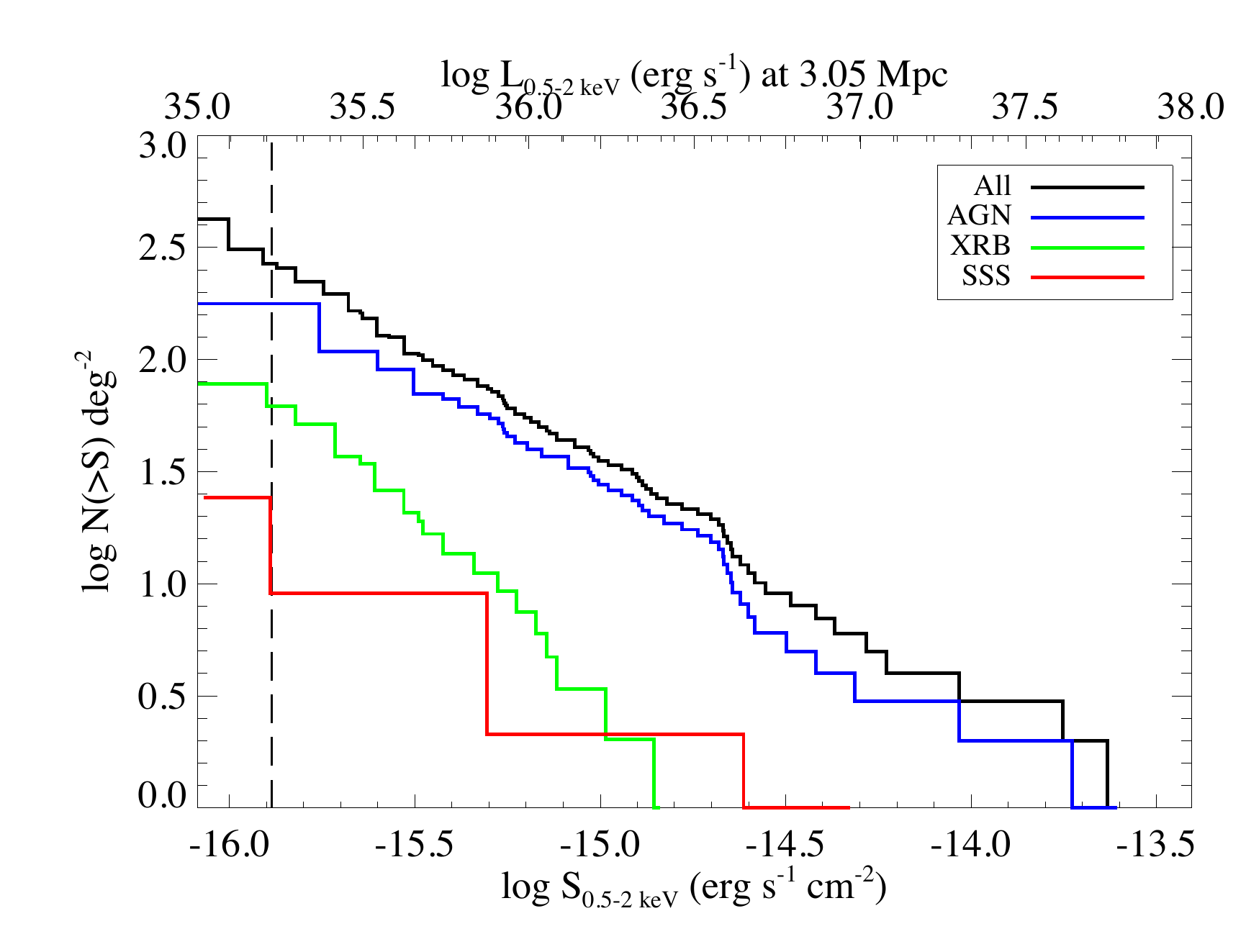}  &
\includegraphics[width=0.5\linewidth,angle=0,clip=true,trim=2cm 12.5cm 2cm 2cm]{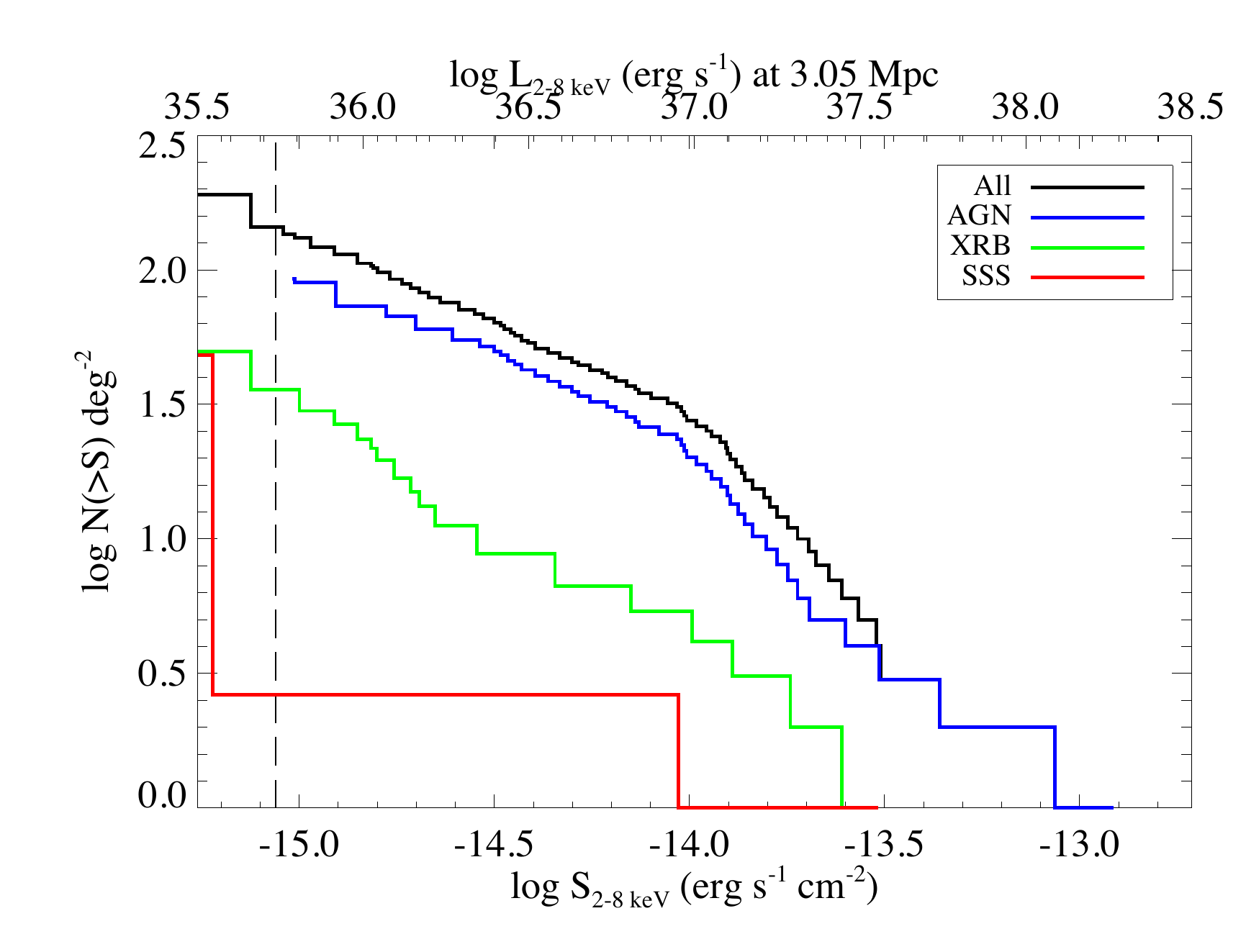}  \\
\end{tabular}
\caption{The cumulative 0.5-2 keV (left) and 2-8 keV (right) X-ray \lognlogs distributions for NGC~404. The dashed line indicates the 90\% completeness limit for each energy band. The \lognlogs distribution for the total X-ray source catalog is shown in black. The \lognlogs distributions evaluated for our AGN candidates, XRB candidates, and SSS candidates are shown in blue, green, and red, respectively.}
\label{XLF}
\end{figure*}

We calculate the cumulative \lognlogs distributions for all X-ray sources in the NGC~404 catalog in the soft (0.5-2 keV) and hard (2-8 keV) bands. Using our X-ray source classifications from Table~\ref{classifications}, we then construct XLFs for those sources designated XRBs, AGNs, and SSSs. We note that the uncertainties in our X-ray source classification scheme will dominate the errors in our \lognlogs fits (discussed more below). The cumulative \lognlogs distributions for the different source categories are shown in Figure~\ref{XLF}.

The two most significant sources of uncertainty in the XRB \lognlogs distributions are the uncertainty in the survey area over which a source could be detected (the $A(S)$) and possible misclassifications of XRBs. For example, based on our survey area and flux range, the \cite{Cappelluti+09} \lognlogs distribution of AGN predicts $\sim$37-40 X-ray sources detected in the NGC~404 catalog are background AGN, implying $\sim$34 X-ray sources should be XRBs. However, we only classify 21 sources as XRB candidates. This discrepancy is the result of the difficulty in separating LMXBs from AGNs, which both exhibit similar X-ray and optical properties. Without further observations (for example, using the angular resolution of \HST to resolve the disks of background galaxies), it can be nearly impossible to differentiate between LMXBs and background AGNs. Despite these uncertainties, we note that the hard \lognlogs distribution of the AGN candidates shows a break near $10^{-14}$ \flux, broadly consistent with the AGN \lognlogs found by both \cite{Cappelluti+09} the \Chandra Deep Field North survey \citep{Luo+08}.

\begin{figure*}
\centering
\begin{tabular}{cc}
\includegraphics[width=0.5\linewidth,angle=0,clip=true,trim=2cm 12.5cm 2cm 2cm]{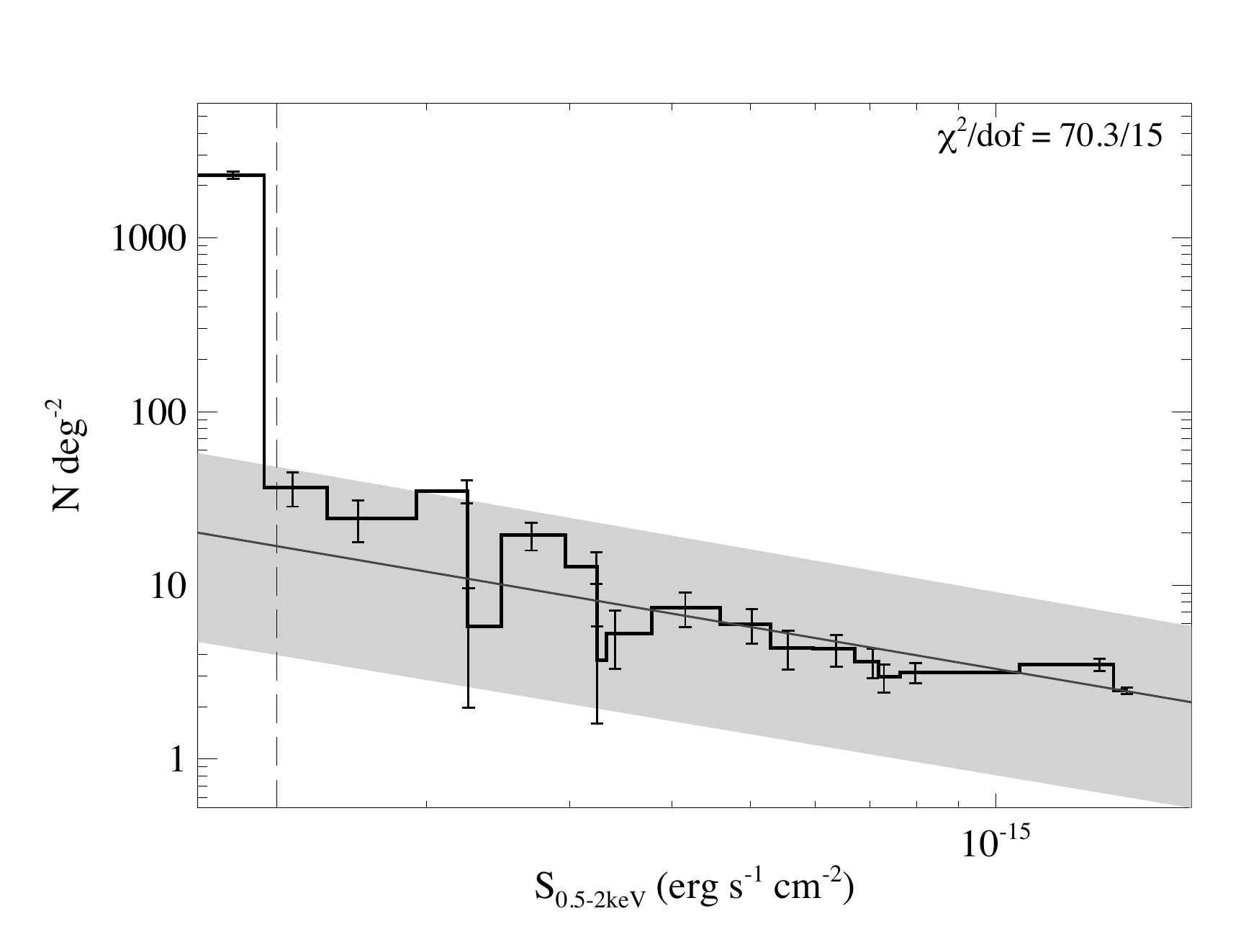}  &
\includegraphics[width=0.5\linewidth,angle=0,clip=true,trim=2cm 12.5cm 2cm 2cm]{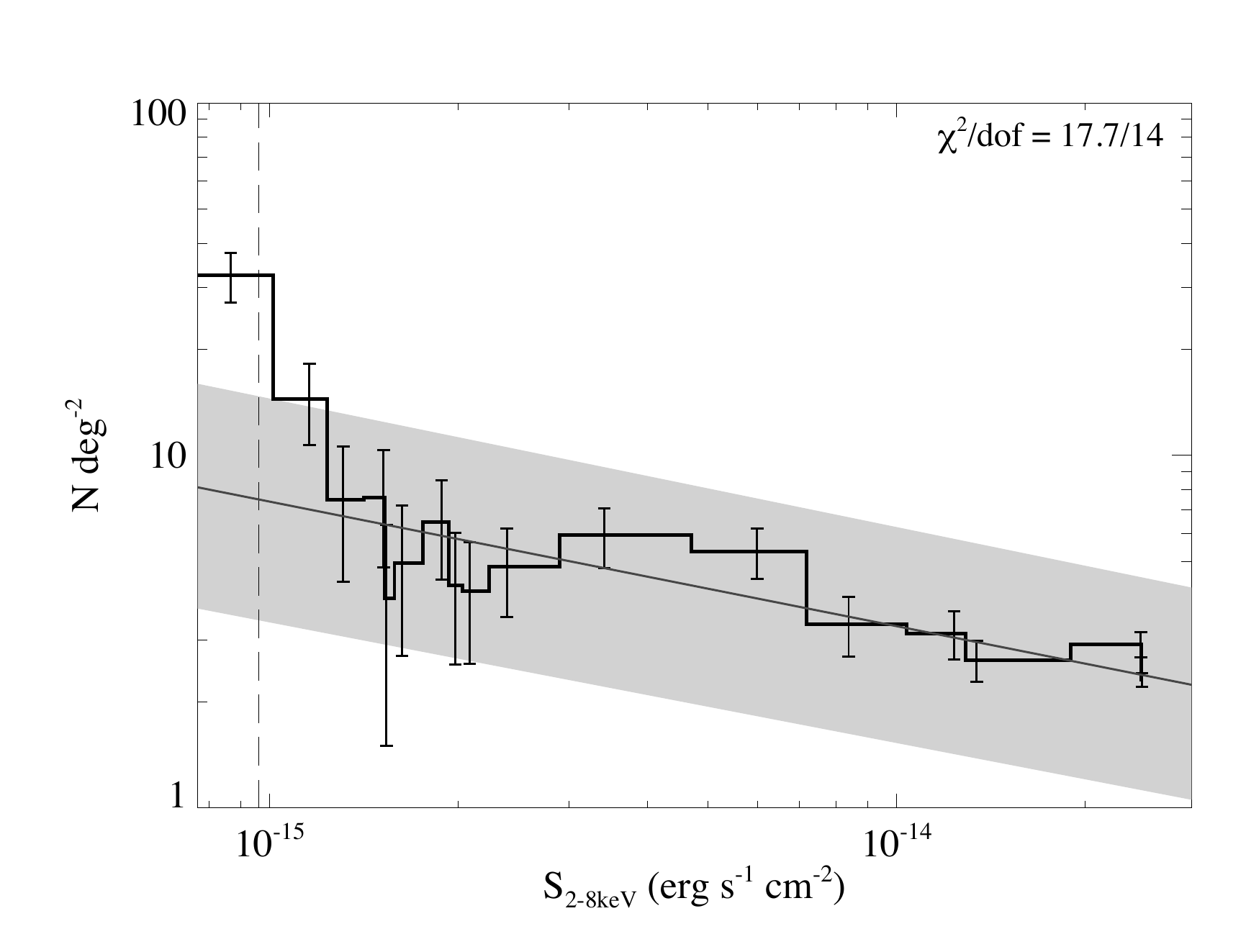}  \\
\includegraphics[width=0.5\linewidth,angle=0,clip=true,trim=2cm 12.5cm 2cm 2cm]{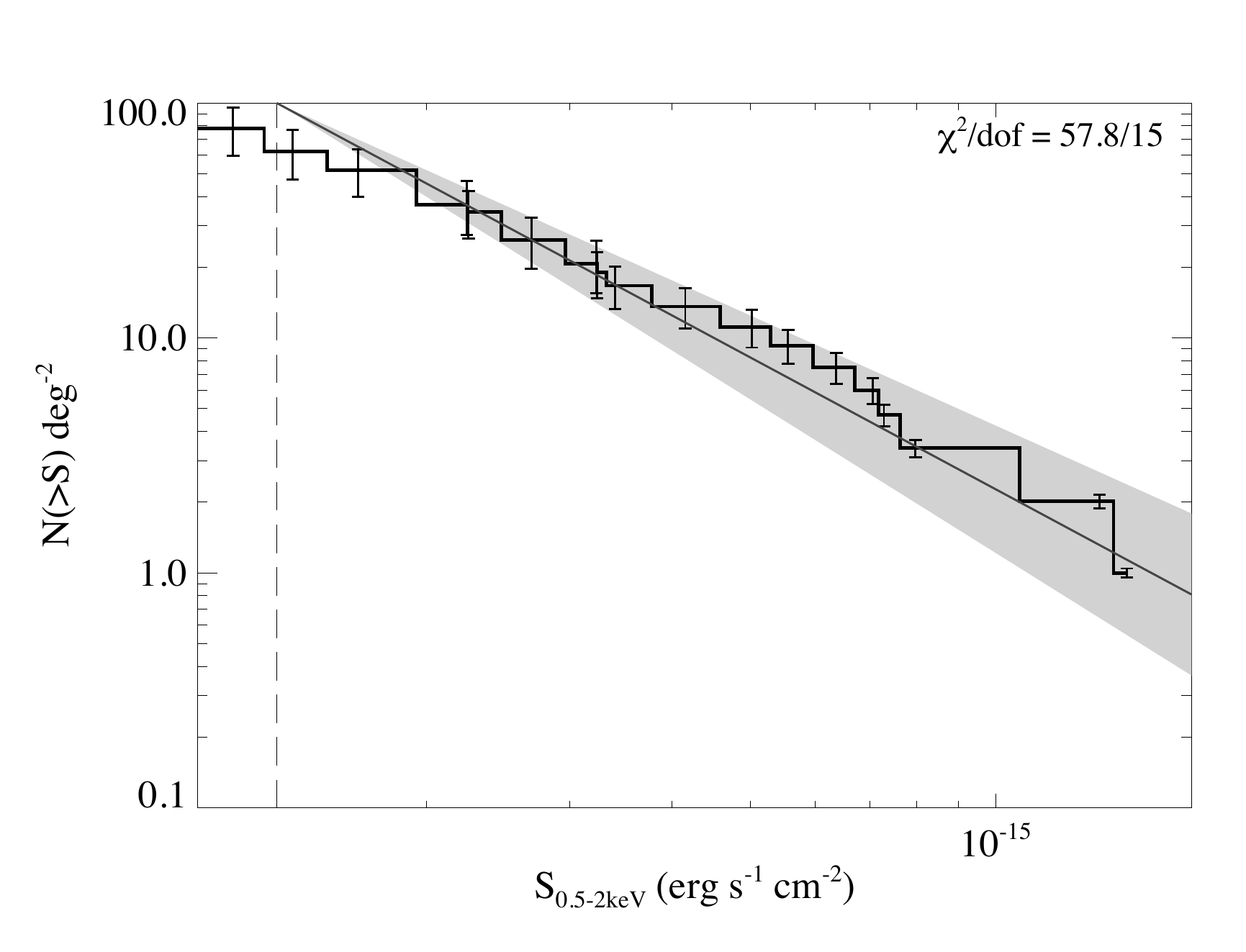}  &
\includegraphics[width=0.5\linewidth,angle=0,clip=true,trim=2cm 12.5cm 2cm 2cm]{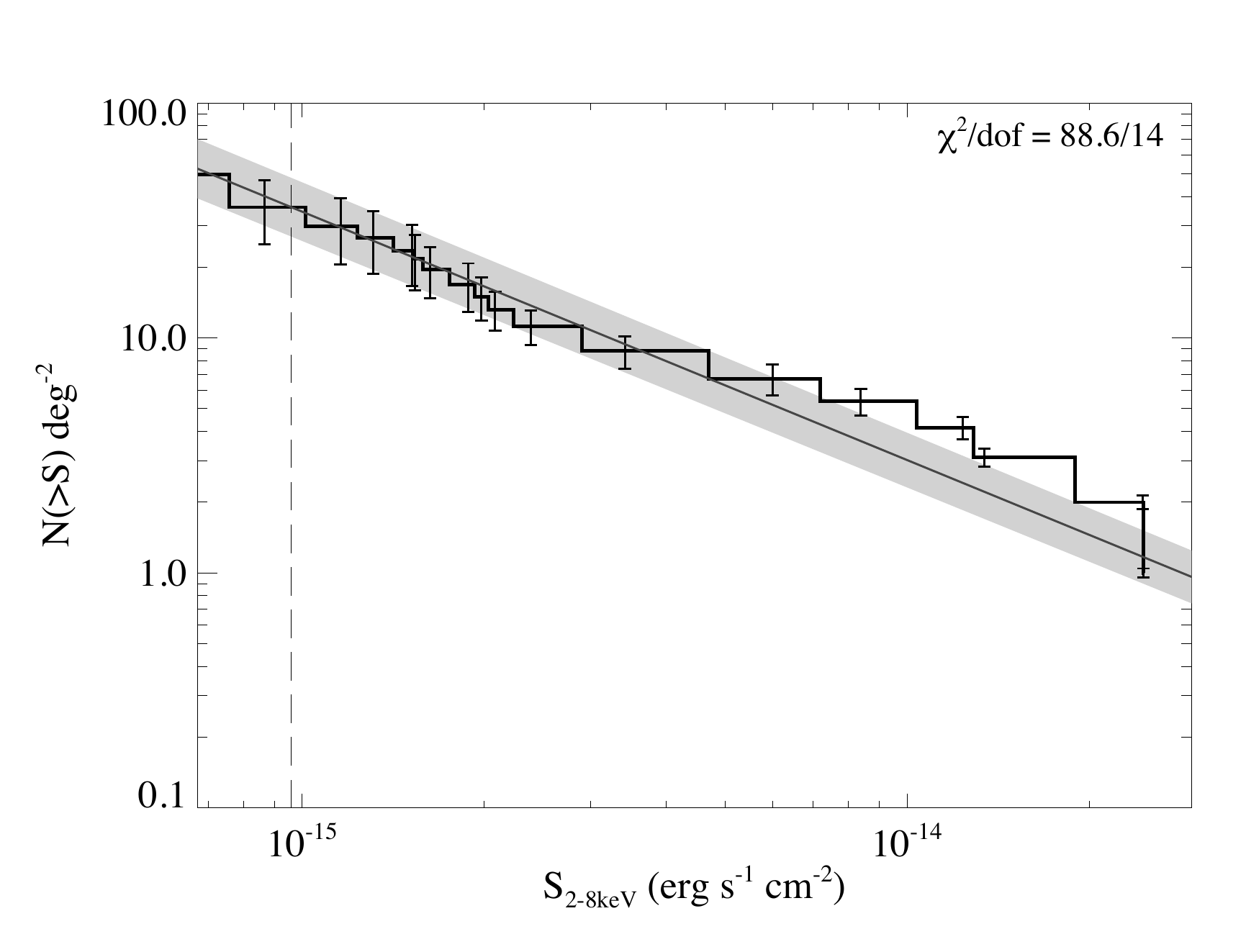}  \\
\end{tabular}
\caption{The top row shows the differential \lognlogs distributions of the NGC~404 XRB candidates in the 0.5-2 keV (left) and 2-8 keV (right) bands, while the bottom row shows the cumulative \lognlogs distributions (again, the soft band is on the left and the hard band on the right). In all four panels, the best-fit power law model is superimposed, with the gray shaded region showing the 1$\sigma$ uncertainty in the fit. The vertical dashed line shows the 90\% limiting flux.}
\label{XLF_models}
\end{figure*}

Both the differential and cumulative XRB \lognlogs distributions in the soft and hard bands are modeled as power laws with the form $N = KS^{-\alpha}$, where $\alpha$ is the power law index and $K$ is a normalization constant, using the IDL routine \texttt{mpfitfun}. The fitting is performed using only the portion of the \lognlogs distribution above the 90\% limiting luminosity. These distributions, along with the best-fit models, are shown in Figure~\ref{XLF_models}. The soft 0.5-2 keV distributions have a best-fit differential power law index of 0.80$\pm$0.14 (\Xdof = 70.3/15) and a cumulative power law index of 1.86$\pm$0.09 (\Xdof = 57.8/15). The hard, 2-8 keV distributions have differential and cumulative power law indices of 0.35$\pm$0.05 (\Xdof = 17.7/14) and 1.09$\pm$0.03 (\Xdof=88.6/14), respectively. 

\begin{figure*}
\centering
\begin{tabular}{cc}
\includegraphics[width=0.5\linewidth,angle=0,clip=true,trim=2cm 12.5cm 2cm 2cm]{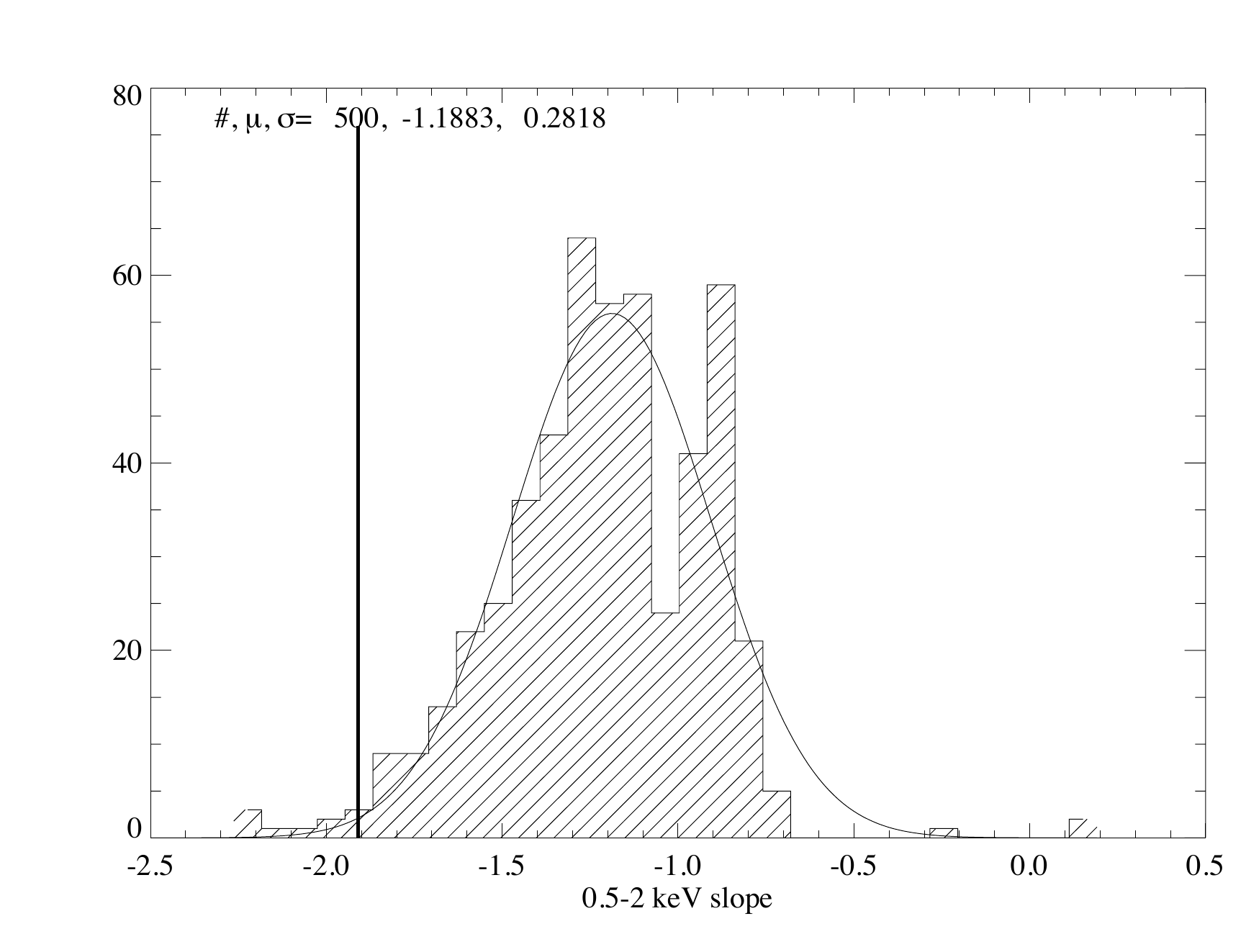}  &
\includegraphics[width=0.5\linewidth,angle=0,clip=true,trim=2cm 12.5cm 2cm 2cm]{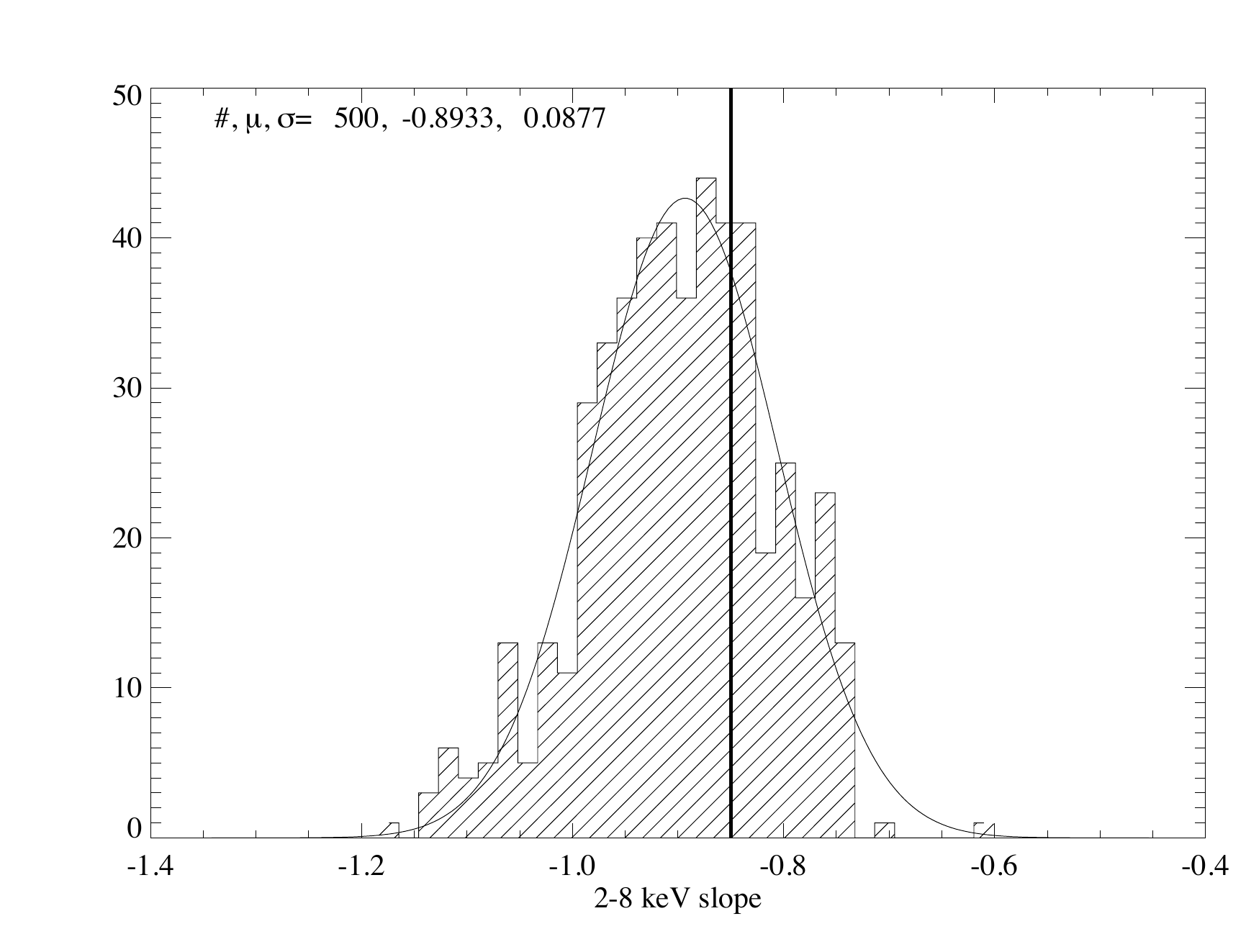}  \\
\end{tabular}
\caption{The distribution of XLF power law slopes for the soft band (left) and the hard band (right) when ten random AGN are included in the XRB \lognlogs distribution. A Gaussian was fit to each of the histogram distributions of best-fit slopes. The thick vertical lines show the best-fit \lognlogs slopes without the the randomly added AGN. The text at the top of each plot shows the number of runs (500 for both energy bands), the mean value of the slopes, and the standard deviation.}
\label{slope_dist}
\end{figure*}

To evaluate the effects of the AGN/LMXB classification ambiguity on our XRB \lognlogs distributions, we randomly select 10 X-ray sources classified as AGNs and add them to our list of XRB candidates. We then recomputed the \lognlogs distributions in both the soft and hard band. This was done 500 times, resulting in 500 \lognlogs distributions for each of the energy bands considered. Using these new \lognlogs distributions, we checked whether (1) a simple power-law model provided an acceptable fit to the new data and (2) if the slope of the best-fit power law model changed when additional sources were included. A single power-law model provided a statistically acceptable fit a clear majority of the 0.5-2 keV \lognlogs distributions. Of the 500 runs, less then $\sim8$\% of the resulting distributions yielded \Xdof$>$1.5, although the typical power law index flattened significantly to $1.19\pm$0.28 (see Figure~\ref{slope_dist}a). While none of the XRB candidates have a 0.5-2 keV flux above $\sim$1.6$\times10^{-15}$ \flux, several AGN candidates do; therefore, misclassifying even a small number of relatively bright LMXBs as AGNs dramatically influences the slope of the soft XLF. 

When the 500 hard \lognlogs distributions were modeled with single and broken power laws, $\sim$52\% of the new 2-8 keV \lognlogs distributions were not well-described by the simple power law model. A broken power law was required to obtain a statistically acceptable fit, with an average break flux of (1.1$\pm$0.3)$\times10^{-14}$ \flux, corresponding to a luminosity of (1.2$\pm$0.4)$\times10^{37}$ \lum at the distance of NGC~404. The typical faint-end slope was 0.78$\pm$0.06 and the typical bright-end slope was 1.50$\pm$0.18. Similar breaks in the LMXB XLF have been reported for other external galaxies \citep[see e.g., ][]{Primini+93,Gilfanov04,Voss+06,Voss+07,Voss+09}. While we do not observe an obvious break in the hard \lognlogs distribution of NGC~404 sources classified as XRBs, even a small number of misclassified AGN could have a significant effect on the shape of the XLF. 

\subsection{Comparison to Other XRB Populations}
The shape of the hard XLF is consistent with observations of LMXB populations (over similar energy ranges) in external galaxies \citep{Primini+93,Gilfanov04,Voss+06,Voss+07,Voss+09} and the Milky Way \citep{Grimm+02,Gilfanov04}. We additionally note that nearly all elliptical galaxies used to construct XLFs in the literature have a larger stellar mass, by an order of magnitude or more, than the dwarf galaxy NGC~404; e.g., Cen A \citep{Wernli+02}, NGC~1407 \citep{Kim+04, Trentham+06}, NGC~4372 \citep{Napolitano+11}. No obvious correlation between XLF slope and stellar mass of the host galaxy is observed.

Several explanations for the $\sim10^{37}$ \lum break in the XLF of LMXBs have been proposed. One particularly attractive explanation involves the nature of the donor companion star: for a persistent, sub-Eddington LMXB, the accretion luminosity $L_X$ is directly proportional to the mass transfer rate \mdot from the secondary star -- thus, mass transfer in a LMXB can be driven by either the loss of orbital angular momentum \citep{Paczynski+81,Verbunt+81} or the increasing radius of the donor due to nuclear evolution \citep{Taam83,Ritter99}. Recently, \cite{Revn+11} showed that LMXB systems with X-ray luminosities below $\sim2\times10^{37}$ \lum have unevolved, main sequence (MS) secondary companions, while systems exhibiting higher X-ray luminosities possess predominantly giant donors. Mass transfer from a giant companion greatly shortens the lifetime of the binary, thus steepening the XLF at the high-luminosity end. The boundary between MS and giant companions has also been found in numerical simulations of LMXB populations by \cite{Fragos+08} and \cite{Kim+09}. 

Unlike its more massive counterparts, NGC~404 experienced a period of enhanced star formation $\sim$0.5 Gyr ago \citep{Williams+10}. Although massive stars would have expired after $\sim$100 Myrs, F and G-type stars formed during this epoch would still be on the MS. This scenario additionally explains the lack of optical counterparts observed with the overlapping \HST exposures -- at a distance of 3.05 Mpc, A-type stars and later would fall below our optical detection limits. Further FUV observations performed by {\it GALEX} suggest that NGC~404 may host a small number of stars as young as $\sim$160 Myr old. Stellar population studies have shown that dwarf ellipticals exhibit, on average, younger stellar ages compared to their massive counterparts \citep{Michielsen+08} -- the same may be true for the X-ray populations of dwarf ellipticals. It is therefore plausible that NGC~404 hosts a population of LMXBs powered by MS donors, not typically found in more massive ellipticals.

The recent ($<$400 Myr) star formation rate (SFR) density of NGC~404 was estimated using \HST observations to be $\sim2\times10^{-4}$ \Msun yr$^{-1}$ kpc$^{-2}$ \citep{Williams+10}. No radial trends in the SFR out to $\sim$5 kpc were observed, implying a typical SFR of $\sim$0.005 \Msun yr$^{-1}$. \cite{Thilker+10} additionally estimated the SFR in the \ion{H}{1} ring at 1.6-6.4 kpc to be 0.0025 \Msun yr$^{-1}$. X-ray emission from HMXBs has been shown to trace recent star formation of the host galaxy \citep{Grimm+03}, and the HMXBs identified in Paper I were successfully used to confirm the SFR of NGC~300. However, the recent SFR of NGC~404 is a factor of $\sim$70 lower than is observed in NGC~300. Using the estimated 2-10 keV luminosity of all of our XRB candidates, we place an upper limit on the NGC~404 SFR of 0.01 \Msun yr$^{-1}$, a factor of $\sim$2-4 larger than the observed rate. This represents one of the first attempts to use X-ray emission to constrain the SFR of an early-type dwarf galaxy.

\section{Summary and Conclusions}\label{end}
We have presented X-ray point source catalogs for a new, deep \Chandra observation of NGC~404 as part of the CLVS. A total of 74 sources were detected in NGC~404 down to a 90\% limiting 0.35-8 keV luminosity of $\sim6\times10^{35}$ \lum if we adopt our standard power law model with \PL=1.9. We evaluated hardness ratios for each X-ray source, performed spectral fitting for those sources with greater than 50 counts, and presented the radial source distribution of the NGC~404 X-ray sources.

We searched for optical counterpart candidates using \HST; 11 X-ray sources were observed in an overlapping \HST field, and only two X-ray sources have candidate optical counterparts. We additionally cross-correlated the X-ray source catalog with {\it GALEX} point source catalog, USNO-B1.0, the 2MASS All-Sky Survey, and the NVSS all-sky radio survey; 31 X-ray sources have at least one multi-wavelength counterpart detected. By combining the results of our X-ray analysis with public multi-wavelength data, we assigned each X-ray source a likely classification (XRB, background AGN, or foreground star).

The differential and cumulative \lognlogs distributions were presented for both the 0.5-2 keV and 2-8 keV bands.  While we do not observe an obvious break in the hard \lognlogs distribution of NGC~404 sources classified as XRBs (as has been observed in other galaxies with X-ray populations dominated by LMXBs), even a small number of misclassified AGN could have a significant effect on the shape of the XLF. We derive the expected ratio of LMXBs to HMXBs in NGC~404. We expect LMXBs to outnumber HMXBs by 4/10/18 at luminosities greater than 10$^{35}$/10$^{36}$/10$^{37}$ \lum.  Additionally, the NGC~404 hosts fewer luminous ($>10^{37}$ \lum) XRBs per stellar mass than its more massive counterparts. The prevalence of low-luminosity XRBs in NGC~404 may be explained by a population of young LMXBs powered by MS companions; unlike its more massive counterparts, NGC~404 is known to have experienced an epoch of enhanced star formation activity $\sim$0.5 Gyr ago. F and G-type stars formed during this epoch would still be on the MS, and would be below our \HST observational detection limits -- thereby also explaining the relative lack of optical counterparts found. 

\acknowledgements
We thank the anonymous referee for numerous detailed and thoughtful comments and suggestions. We also thank the editor, Eric Feigelson, for helpful comments. Support for this work was provided by the National Aeronautics and Space Administration through Chandra Award Number G01-12118X issued by the Chandra X-ray Observatory Center, which is operated by the Smithsonian Astrophysical Observatory for and on behalf of the National Aeronautics Space Administration under contract NAS8-03060. T.J.G acknowledges support under NASA contract NAS8-03060. This research has made use of the NASA/IPAC Extragalactic Database (NED) which is operated by the Jet Propulsion Laboratory, California Institute of Technology, under contract with the National Aeronautics and Space Administration.

\bibliography{apjmnemonic,NGC404}
\bibliographystyle{apj}
                                                                                                    
\end{document}